%% file: 01.Eloss.tex
\documentclass[onecolumn, prd, aps, tightenlines, preprintnumbers, showpacs, nofootinbib, superscriptaddress, notitlepage]{revtex4-1}

\pdfoutput=1

\usepackage{float}
\usepackage{amsmath}
\usepackage{color}
\usepackage{graphicx}
\usepackage[dvipsnames]{xcolor}
\usepackage{url}
\usepackage{epsfig}
\usepackage[T1]{fontenc}
\usepackage{multirow}
\usepackage{physics} 
\usepackage{booktabs} 
\usepackage{array} 
\usepackage{paralist} 
\usepackage{grffile}
\usepackage{verbatim} 
\usepackage{subfig} 
\usepackage{amsmath,amsthm,amssymb,bm,amsfonts}
\usepackage{slashed}
\usepackage[utf8]{inputenc}
\usepackage{hyperref}

\definecolor{darkpastelgreen}{rgb}{0.01, 0.55, 0.24}

\usepackage{color}
\usepackage[normalem]{ulem}

\begin{document}

\title{Coherent energy loss effects in dihadron azimuthal angular 
  correlations in Deep Inelastic Scattering at small $x$}

\author{Filip Bergabo}
\email{fbergabo@gradcenter.cuny.edu}
\affiliation{Department of Natural Sciences, Baruch College, CUNY, 17 Lexington Avenue, New York, NY 10010, USA}
\affiliation{City University of New York Graduate Center, 365 Fifth Avenue, New York, NY 10016, USA}

\author{Jamal Jalilian-Marian}
\email{jamal.jalilian-marian@baruch.cuny.edu}
\affiliation{Department of Natural Sciences, Baruch College, CUNY, 17 Lexington Avenue, New York, NY 10010, USA}
\affiliation{City University of New York Graduate Center, 365 Fifth Avenue, New York, NY 10016, USA}


\begin{abstract}
We perform an exploratory study of the role of coherent,  
medium-induced energy loss in azimuthal angular correlations in dihadron 
production in Deep Inelastic Scattering (DIS) at small $x$ where the
target proton/nucleus is modeled as a Color Glass Condensate. In this 
approach coherent radiative energy loss is part of the higher order corrections
to the leading order dihadron production cross section. We include
the effects of both gluon saturation and coherent radiative energy loss 
and show that radiative cold-matter energy loss has a significant effect 
on the so-called coincidence probability for the back to back production 
of dihadrons in DIS. We also define a double ratio of coincidence probabilities
for a nucleus and proton targets and show that it is very robust against 
higher order radiative corrections.

\end{abstract}

\maketitle


\section{Introduction}
The rise of parton (and especially gluon) distribution functions of a 
proton with decreasing Bjorken $x$ as observed in Deep Inelastic 
Scattering (DIS) experiments at HERA~\cite{Aaron:2009kv} was a pleasant 
surprise which triggered intense theoretical and experimental studies 
of the behavior of QCD scattering cross sections at small $x$ (equivalently, 
at high energy). This observed rise of the gluon distribution function
can not however go on forever and must be tamed by high gluon density
effects, the so-called gluon saturation~\cite{Gribov:1984tu,Mueller:1985wy}. The color 
glass condensate (CGC) formalism~\cite{Gelis:2010nm} is an effective theory 
of high energy (or equivalently small $x$) QCD which includes gluon 
saturation effects. In this formalism the small $x$ gluon modes of a 
fast-moving proton or nucleus are collectively represented as a 
classical color field generated by the large $x$ color degrees of freedom 
treated as static color charges~\cite{McLerran:1993ni,Jalilian-Marian:1996mkd}. A high energy 
collision involving two hadrons/nuclei at small $x$ in this approach 
is thus treated as a collision of two classical color fields, i.e. 
two color shock waves. On the other hand in DIS at small $x$ one 
has a two-stage process where the virtual photon first splits into a 
quark anti-quark pair (a dipole) which subsequently scatter from the 
target hadron/nucleus modeled as a classical color field. Higher order 
loop corrections then lead to energy (equivalently $x$ or rapidity)
dependence of the quark anti-quark dipole-hadron/nucleus scattering 
cross section.

There have been numerous applications of the CGC formalism to particle 
production in high energy proton-proton, proton-nucleus and 
nucleus-nucleus collisions as well as to fully inclusive structure 
functions in DIS~\cite{Albacete:2017qng}. While there are strong 
hints for the presence 
of significant saturation effects in particle production spectra 
in the high energy hadronic and nuclear collisions at 
RHIC~\cite{Aschenauer:2016our} and the LHC, more differential 
measurements in a cleaner environment~\cite{Accardi:2012qut} and higher 
precision theoretical calculations are needed to clearly establish 
gluon saturation as the dominant dynamics in the observed particle 
spectra at small $x$. 

Two-particle production and azimuthal angular correlations are perhaps 
the most sensitive probe of saturation dynamics and as such have been 
intensively studied in the 
CGC formalism~\cite{JalilianMarian:2004da,Jalilian-Marian:2005qbq, Marquet:2007vb,Albacete:2010pg,Stasto:2011ru,Lappi:2012nh,Jalilian-Marian:2012wwi,Jalilian-Marian:2011tvq, Zheng:2014vka,Stasto:2018rci,Albacete:2018ruq,Mantysaari:2019hkq,Hatta:2020bgy,Jia:2019qbl}. 
The Color Glass Condensate formalism 
predicts a broadening and eventual disappearing of the away side peak 
in dihadron back-to-back correlations~\cite{Marquet:2007vb} as 
experimentally observed in forward rapidity proton (deuteron)-nucleus 
collisions at RHIC~\cite{Braidot:2010ig,Adare:2011sc}. While Leading 
Order CGC calculations of dihadron production and angular correlations  
with (or without) running coupling corrections describe the experimental 
data quite well there may be other effects which also significantly 
contribute to this disappearance of the away side peak, for example, 
cold matter energy loss where one of the produced partons scatters from 
the nuclear target and radiates away some of 
its energy. Indeed it has been shown~\cite{Kang:2011bp} that combining 
phenomenologically-motivated models of cold matter energy loss with 
models of nuclear shadowing of parton distribution functions can also 
describe the experimental data. Therefore it is prudent to understand 
how important cold matter energy loss effects are as compared 
with gluon saturation. It should be noted that the coherent energy loss
as we define here is part of the Next to Leading Order (NLO) corrections
to the Leading Order (LO) dihadron production cross section. 
In the Color Glass Condensate formalism this is true to any order in the 
coupling constant where the coherent energy loss is a higher order in 
$\alpha_s$ correction to a fixed order calculation. 
Nevertheless as NLO corrections to this process~\cite{nlo:fbjjm} are not 
currently known~\footnote{As this manuscript was being finalized we became 
aware of a very recent NLO calculation of dijet production~\cite{Caucal:2021ent}.} it is therefore useful   
to have a quantitative estimate of coherent energy loss effects on the 
dihadron azimuthal angular correlations.

In this exploratory work we study dihadron (quark anti-quark) 
azimuthal angular correlations in the back-to-back kinematics 
in DIS at small $x$ where both gluon saturation and coherent 
cold matter energy loss are included using the same formalism. 
First, we re-derive the cross section for production of a quark, 
anti-quark and a gluon in DIS which was already done 
in~\cite{Ayala:2016lhd,Ayala:2017rmh}. 
We then take the soft gluon limit and integrate over the final 
state gluon transverse momentum and compare the soft gluon 
radiation spectra, normalized to no radiation, between a nucleus and a
proton target. Gaussian approximation is used to calculate the
correlation functions of Wilson lines appearing as dipoles and 
quadrupoles which efficiently contain all the target information.
We show that medium-induced coherent energy loss is most significant 
at the back to back limit and drops off as one goes away from this limit, and as one considers higher photon virtualities.  
We then consider the contribution of coherent energy loss to the away 
side peak in dihadron correlations in the back-to-back 
kinematics and show that it is significant. We then define a double ratio of coincidence 
probabilities and show that this double ratio is very robust against
NLO corrections.
We finish by outlining the steps needed for a more realistic study of 
medium-induced energy loss effects in dihadron angular correlations.


\section{Coherent energy loss in DIS at small $x$ from CGC}
\label{sec2}
The leading order process for dihadron~\footnote{Everywhere 
in this paper we will consider partons rather than hadrons in
the final state.} production in DIS at small 
$x$ is the splitting of the virtual photon into a quark anti-quark 
pair which then multiply scatters on the target proton or nucleus. 
In the eikonal approximation inherent at small $x$ it is assumed that
the energy of the photon, and hence of the quark anti-quark pair,
is so large that their recoil can be neglected and the pair stays
on straight line trajectories while passing through the target. The
scattering amplitude contains two Wilson lines~\cite{Gelis:2002nn} 
(multiple scatterings of each parton from the target is re-summed 
into a Wilson line) so that dihadron production cross section 
involves not only dipoles but also quadrupoles, correlation functions 
of two and four Wilson lines. These dipoles and quadrupoles satisfy the 
BK/JIMWLK evolution equation~\cite{Balitsky:1995ub,Kovchegov:1999ua,Jalilian-Marian:1997qno,Jalilian-Marian:1997jhx,Jalilian-Marian:1997ubg,Kovner:2000pt,Iancu:2000hn,Ferreiro:2001qy} which governs their energy (rapidity or $x$) 
dependence~\cite{Dumitru:2011vk,Dumitru:2011zz,Dumitru:2010ak}. In the Color Glass Condensate formalism 
multiple scatterings and rapidity evolution result in the broadening 
and reduction of the away side peak in dihadron azimuthal angular 
correlations.

As either quark or anti-quark radiates a gluon, the energy carried away
by the not-measured soft gluon will look as if it is lost in the process.
Following Munier, Peign\'e and Petreska~\cite{Munier:2016oih} we define 
the medium-induced radiation spectrum as 

\begin{align}
z_3 \frac{ \dd I}{\dd z_3} = \frac{ \frac{ \dd \sigma^{\gamma^* A \to q\bar{q} g X}}{\dd^2 \mathbf{p}\, \dd^2 \mathbf{q} \, \dd y_1 \, \dd y_2 \, \dd y_3}}{\frac{ \dd \sigma^{\gamma^* A \to q\bar{q} X}}{\dd^2 \mathbf{p} \, \dd^2 \mathbf{q} \, \dd y_1 \, \dd y_2}}. \label{radspec}
\end{align}
where $y_1, y_2, y_3$ are rapidities of the quark, anti-quark and radiated
gluon respectively while $\bf{p}, \bf{q}$ are the transverse 
momenta of the quark and anti-quark and the transverse momentum of the
gluon is integrated over. Here $z_3$ is the radiated gluon's fraction of the photon's plus momentum. We note that the three-parton production 
cross section in DIS at small $x$ is already computed 
in~\cite{Ayala:2016lhd,Ayala:2017rmh}. Integrating over the
gluon momentum gives the contribution of the real corrections 
in the Next to Leading Order corrections to the Leading Order quark 
anti-quark production in DIS at small $x$. The medium-induced 
coherent energy loss is then defined~\cite{Munier:2016oih} 
as the difference in radiation spectra between a nucleus and 
a proton target,
 
\begin{align}
z_3 \frac{ \dd I_{\text{ind}}}{ \dd z_3} \equiv z_3 \frac{ \dd I_A}{\dd z_3} - z_3 \frac{ \dd I_p}{\dd z_3} =  \frac{ \frac{ \dd \sigma^{\gamma^* A \to q\bar{q} g X}}{\dd^2 \mathbf{p}\, \dd^2 \mathbf{q} \, \dd y_1 \, \dd y_2 \, \dd y_3}}{\frac{ \dd \sigma^{\gamma^* A \to q\bar{q} X}}{\dd^2 \mathbf{p} \, \dd^2 \mathbf{q} \, \dd y_1 \, \dd y_2}} -  \frac{ \frac{ \dd \sigma^{\gamma^* p \to q\bar{q} g X}}{\dd^2 \mathbf{p}\, \dd^2 \mathbf{q} \, \dd y_1 \, \dd y_2 \, \dd y_3}}{\frac{ \dd \sigma^{\gamma^* p \to q\bar{q} X}}{\dd^2 \mathbf{p} \, \dd^2 \mathbf{q} \, \dd y_1 \, \dd y_2}}. \label{eloss}
\end{align}
where an integration over the transverse momentum of the radiated gluon
is implied in the numerators. 

Suppression of the away side peak in two-particle correlations as a function of the azimuthal angle $\Delta \phi$ between the two outgoing particles in forward rapidity deuteron-nucleus collisions was predicted in~\cite{Marquet:2007vb} using leading order calculations of the coincidence probability $\text{CP}(\Delta \phi)$,  here defined as~\cite{Albacete:2010pg},

\begin{align}
\text{CP}(\Delta \phi) = \frac{N_{\text{pair}}(\Delta \phi)}{N_{\text{trig}}}, \,\,\,\,\,\, N_{\text{pair}}(\Delta \phi) = \int_{p_{\text{min}}}^{p_{\text{max}}} p\, \dd p \int_{q_{\text{min}}}^{q_{\text{max}}} q \, \dd q\frac{\dd \sigma^{\gamma^* A \to q\bar{q} X}}{\dd^2 \mathbf{p} \, \dd^2 \mathbf{q} \, \dd y_1 \, \dd y_2}, \,\,\,\,\,\, N_{\text{trig}} = \int_{p_{\text{min}}}^{p_{\text{max}}} p\, \dd p \frac{\dd \sigma^{\gamma^*A \to qX}}{\dd^2 \mathbf{p}\, \dd y_1}. \label{CP}
\end{align}
for dihadron production in DIS. $\text{CP}(\Delta \phi)$ is a commonly studied observable which represents the probability per unit angle for correlated production of two hadrons; a leading (trigger) hadron with transverse momentum $|\mathbf{p}|$ between $p_{\text{min}}$ and $p_{\text{max}}$ accompanied by an away side hadron with transverse momentum $|\mathbf{q}|$ between $q_{\text{min}}$ and $q_{\text{max}}$ with an azimuthal angular separation of $\Delta \phi$. We will explore the contribution of fully coherent cold matter energy loss to CP$(\Delta \phi)$ by adding radiative corrections to $N_{\text{pair}}$ while using the leading order result for $N_{\text{trig}}$. In this preliminary study we will assume fixed rapidities to simplify our calculations.

We use the spinor helicity methods to calculate the quark anti-quark 
and quark anti-quark gluon production amplitudes. In the latter case 
there are four diagrams corresponding to the four possibilities when 
the gluon is radiated from either the quark or anti-quark and before 
or after the scattering from the target (see~\cite{Ayala:2017rmh} for 
more details), 
\begin{figure}[H] 
\begin{center}
\includegraphics[width=46mm]{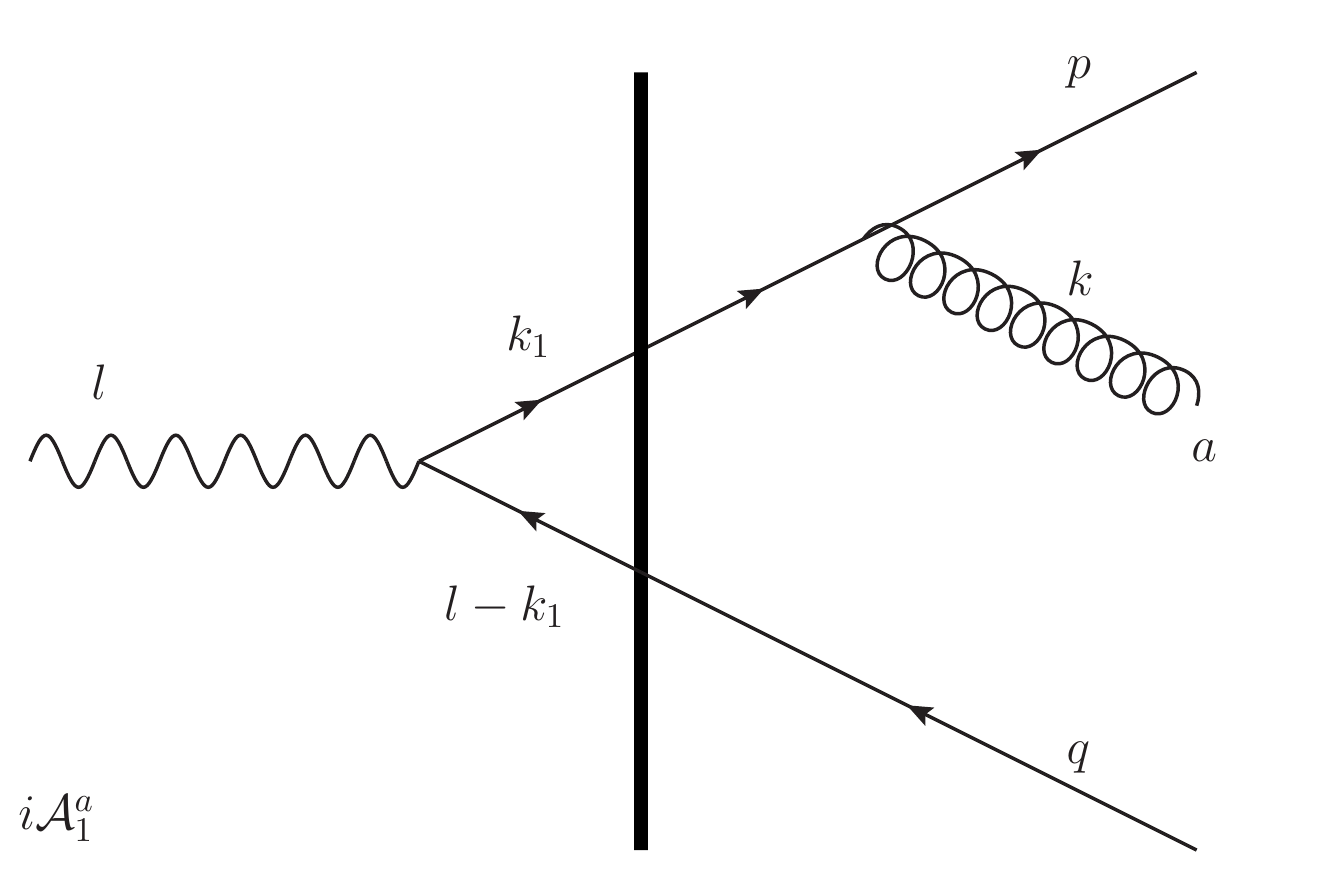}\includegraphics[width=46mm]{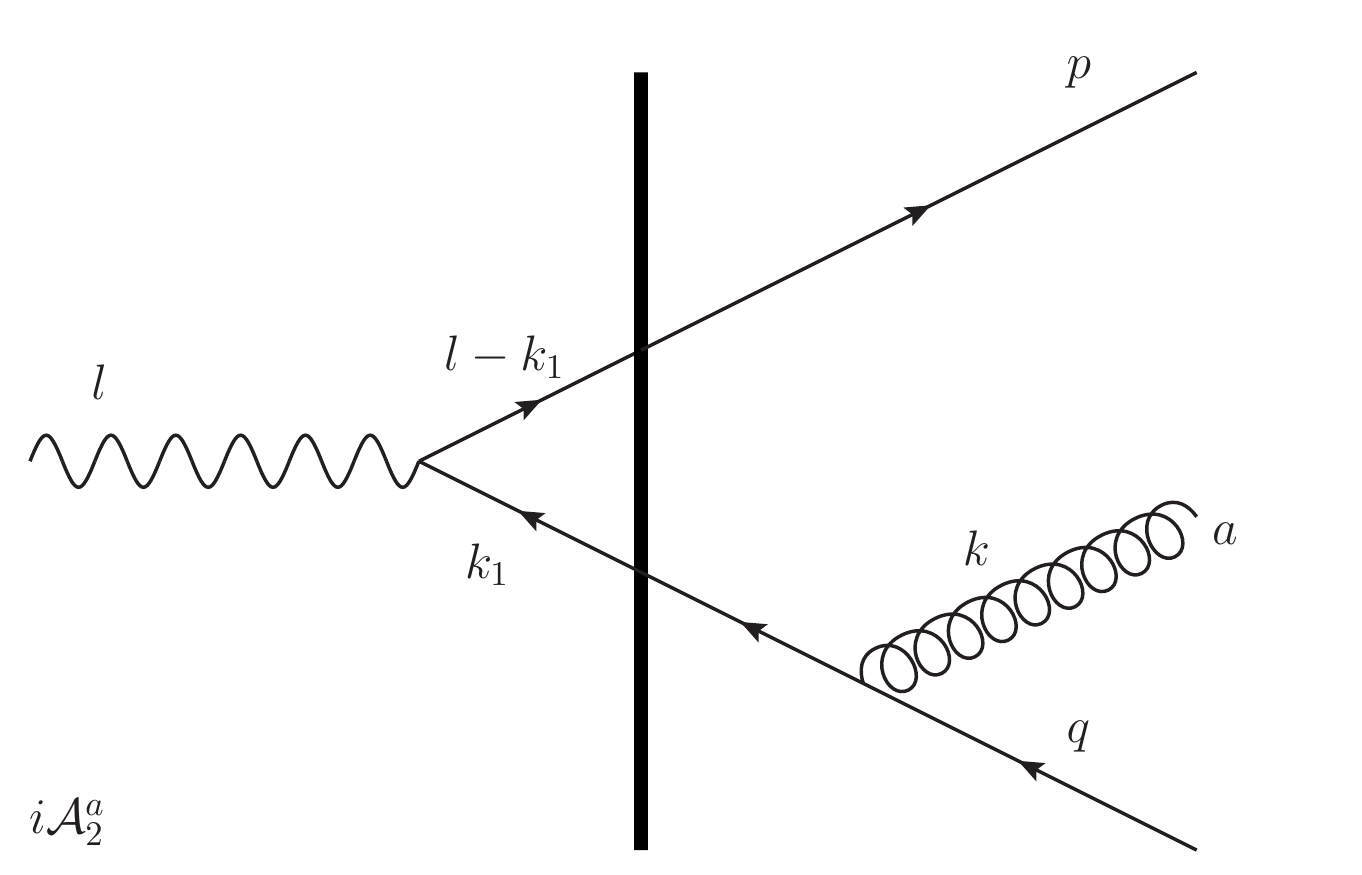}\includegraphics[width=46mm]{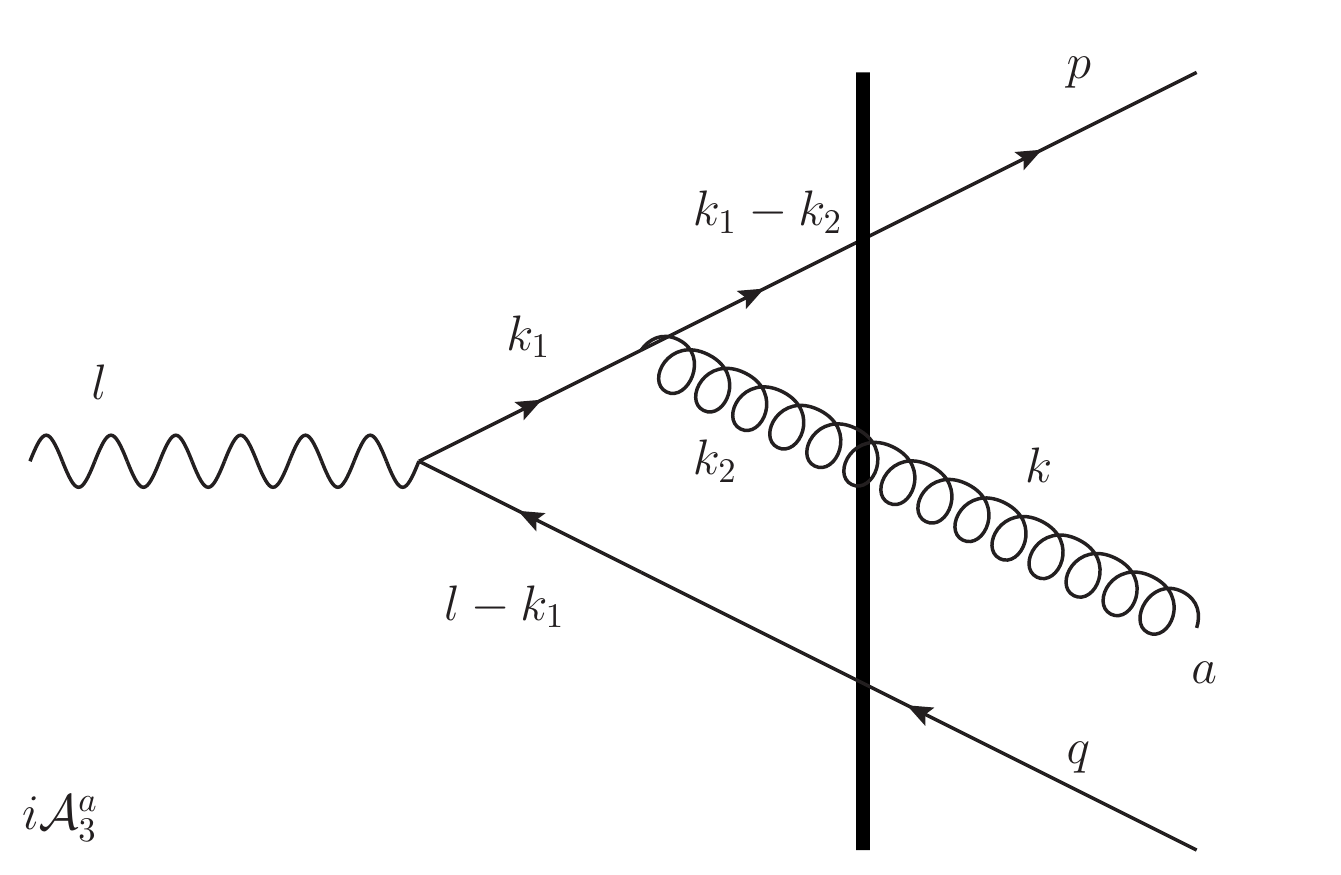}\includegraphics[width=46mm]{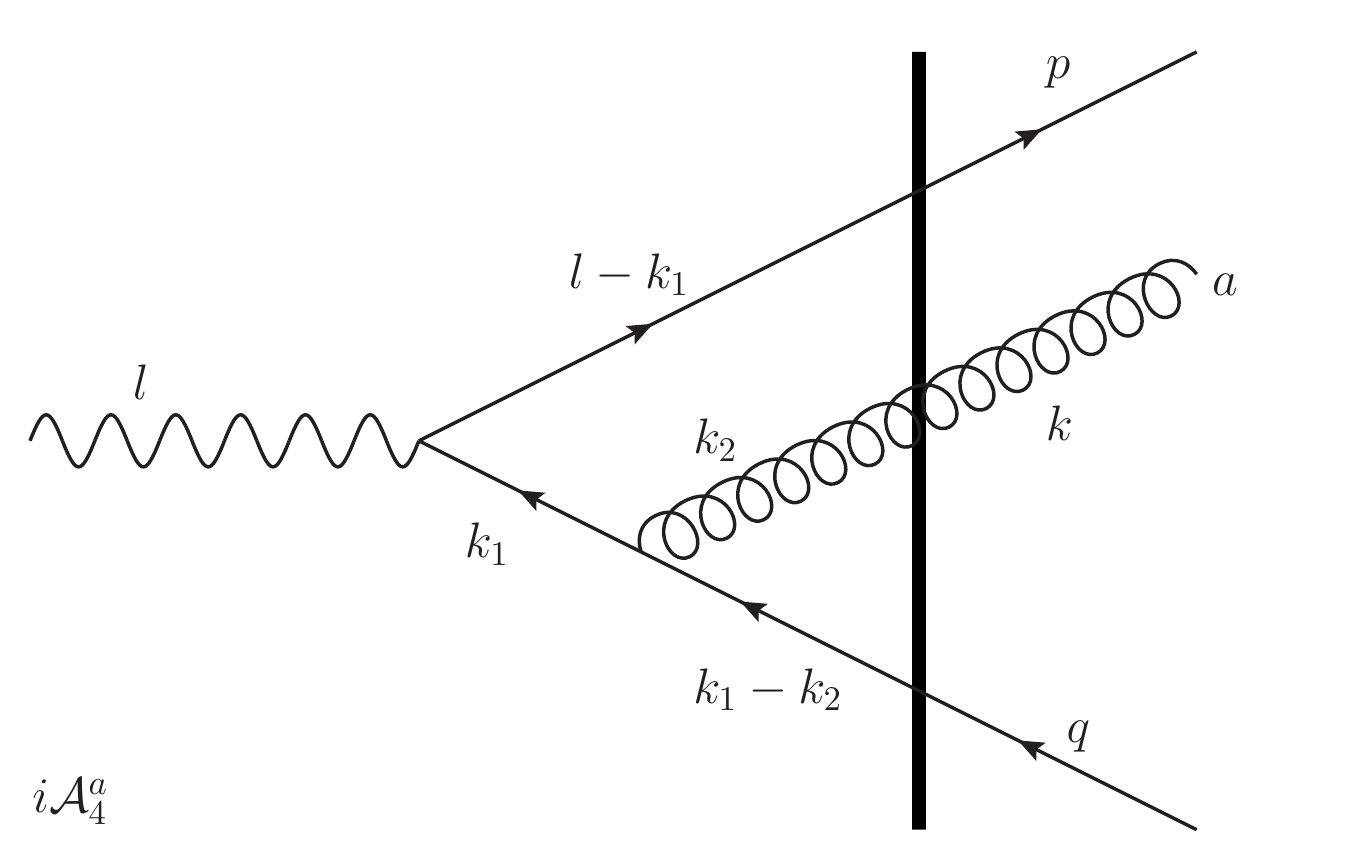}
\end{center}
\caption{The four diagrams for three parton production $\gamma^* A \to q\bar{q}gX$. The solid vertical line represents multiple scatterings from the target.}\label{diagrams}
\end{figure}
\noindent which can be evaluated to give  
\begin{align}
i\mathcal{M}_1^a = &8egl^+ \int \dd^2 \mathbf{x}_1 \dd^2 \mathbf{x}_2 t^a V(\mathbf{x}_1)V^\dag(\mathbf{x}_2) \int \frac{ \dd^2\mathbf{k}_1}{(2\pi)^2} \frac{ N_1 e^{i\mathbf{k}_1\cdot(\mathbf{x}_1-\mathbf{x}_2)} e^{-i(\mathbf{p}+\mathbf{k})\cdot \mathbf{x}_1}  e^{-i\mathbf{q}\cdot \mathbf{x_2}}}{\mathbf{k}_1^2 + z_2(1-z_2)Q^2}.\nonumber \\
i\mathcal{M}_2^a = & 8egl^+ \int \dd^2 \mathbf{x}_1\dd^2 \mathbf{x}_2 V(\mathbf{x}_1)V^\dag(\mathbf{x}_2)t^a  \int \frac{ \dd^2\mathbf{k}_1}{(2\pi)^2} \frac{ N_2 e^{i\mathbf{k}_1\cdot(\mathbf{x}_2-\mathbf{x}_1)} e^{-i(\mathbf{q}+\mathbf{k})\cdot \mathbf{x}_2} e^{-i\mathbf{p} \cdot \mathbf{x}_1}}{\mathbf{k}_1^2 + z_1(1-z_1)Q^2}.\nonumber \\
i\mathcal{M}_3^a =& \frac{8egl^+}{z_1} \int \dd^2 \mathbf{x}_1 \dd^2 \mathbf{x}_2\dd^2\mathbf{x}_3 V(\mathbf{x}_1)t^bV^\dag(\mathbf{x}_2)U(\mathbf{x}_3)^{ba}  \int \frac{\dd^2\mathbf{k}_1}{(2\pi)^2} \frac{ \dd^2 \mathbf{k}_2}{(2\pi)^2}  \frac{ N_3  \,e^{i\mathbf{k}_1\cdot(\mathbf{x}_1-\mathbf{x}_2)}e^{i\mathbf{k}_2\cdot(\mathbf{x}_3-\mathbf{x}_1)} e^{-i\mathbf{k}\cdot\mathbf{x}_3} e^{-i\mathbf{p}\cdot \mathbf{x}_1} e^{-i\mathbf{q}\cdot \mathbf{x}_2}}{\left[\mathbf{k}_1^2 + z_2(1-z_2)Q^2\right] \left[Q^2 + \frac{\mathbf{k}_1^2}{z_2} + \frac{\mathbf{k}_2^2}{z_3} + \frac{(\mathbf{k}_1-\mathbf{k}_2)^2}{z_1}\right]} .\nonumber \\
i\mathcal{M}_4^a =& \frac{8egl^+}{z_2} \int \dd^2 \mathbf{x}_1 \dd^2 \mathbf{x}_2 \dd^2 \mathbf{x}_3  V(\mathbf{x}_1)t^bV^\dag(\mathbf{x}_2)U(\mathbf{x}_3)^{ba}\int \frac{\dd^2 \mathbf{k}_1}{(2\pi)^2} \frac{ \dd^2\mathbf{k}_2}{(2\pi)^2}  \frac{ N_4  \,e^{i\mathbf{k}_1\cdot(\mathbf{x}_2 - \mathbf{x}_1)}e^{i\mathbf{k}_2 \cdot(\mathbf{x}_3 - \mathbf{x}_2)} e^{-i\mathbf{k}\cdot \mathbf{x}_3} e^{-i\mathbf{p}\cdot\mathbf{x}_1} e^{-i\mathbf{q}\cdot\mathbf{x}_2}}{\left[\mathbf{k}_1^2 + z_1(1-z_1)Q^2\right] \left[Q^2 + \frac{\mathbf{k}_1^2}{z_1} + \frac{\mathbf{k}_2^2}{z_3} + \frac{(\mathbf{k}_1-\mathbf{k}_2)^2}{z_2}\right]} .\label{amps}
\end{align}

\noindent where we have factored out the overall momentum ($+$ component only) conserving delta function $2\pi\delta(l^+ - p^+ - q^+ - k^+)$ (not shown and hence the reason for denoting the amplitude as $i \mathcal{M}$ rather than 
$i \mathcal{A}$ as in the figure). We have also defined $z_1, z_2, z_3$ as fractions of the virtual photon energy carried by the quark, anti-quark 
and gluon respectively. $V(\mathbf{x}_i)$ is a Wilson line in the fundamental representation, while $U(\mathbf{x}_i)^{ba}$ is a Wilson line in the adjoint representation. The numerators $N_i$ contain the spinor structures 
of the amplitude and are defined as

\begin{align}
N_1 =\frac{\bar{u}(p) \slashed{\epsilon}^*(k) (\slashed{p} + \slashed{k}) \slashed{n} \slashed{k}_1 \slashed{\epsilon}(l)(\slashed{k}_1 - \slashed{l})\slashed{n} v(q)}{16(l^+)^2(p+k)^2}, \,\,\,\,\,\,\,N_2& = \frac{ \bar{u}(p) \slashed{n} (\slashed{l} - \slashed{k}_1)\slashed{\epsilon}(l) \slashed{k}_1 \slashed{n}(\slashed{q} + \slashed{k}) \slashed{\epsilon}^*(k) v(q)}{16(l^+)^2(q+k)^2}, \nonumber \\
N_3 =\frac{\bar{u}(p) \slashed{n} (\slashed{k}_1 - \slashed{k}_2) \gamma^\mu  \slashed{k}_1 \slashed{\epsilon}(l)(\slashed{l} - \slashed{k}_1) \slashed{n} v(q) d_{\mu \nu}(k_2) \epsilon^\nu(k)^*}{16(l^+)^2}, \,\,\,\,\,\, N_4& =\frac{ \bar{u}(p) \slashed{n} (\slashed{l} - \slashed{k}_1) \slashed{\epsilon}(l) \slashed{k}_1 \gamma^\mu (\slashed{k}_2 - \slashed{k}_1) \slashed{n} v(q) d_{\mu \nu}(k_2) \epsilon^\nu(k)^*}{16(l^+)^2}. \label{numerators}
\end{align}

\noindent In this exploratory work we consider a longitudinally 
polarized photon and use the spinor helicity formalism to evaluate 
these numerators for a given parton helicity. The integrals over 
$\mathbf{k}_1$ and $\mathbf{k}_2$ can then be performed. Here we show 
the amplitudes for the specific case of a positive helicity quark 
(so that anti-quark has negative helicity) and gluon, and a 
longitudinally polarized photon.
\begin{align}
i\mathcal{M}_1^{a\, L:+,+} = &8eg l^+\int \dd^2 \mathbf{x}_1 \dd^2 \mathbf{x}_2 t^a V(\mathbf{x}_1)V^\dag(\mathbf{x}_2)e^{-i\mathbf{p}\cdot\mathbf{x}_1}e^{-i\mathbf{q}\cdot\mathbf{x}_2} \nonumber \\
 & \left[ \frac{ -Q (z_1 z_2)^{\frac32} (1-z_2)}{(2\pi)}\right] K_0\left(|\mathbf{x}_{12}| Q_2\right) \frac{e^{-i\mathbf{k}\cdot\mathbf{x}_1}}{(p+k)^2} \left( \frac{\mathbf{k}\cdot \boldsymbol{\epsilon}}{z_3} - \frac{ \mathbf{p}\cdot \boldsymbol{\epsilon}}{z_1}\right).\label{ia1final2}  \\
i\mathcal{M}_2^{a\, L;+,+} = &8egl^+\int \dd^2\mathbf{x}_1 \dd^2\mathbf{x}_2 V(\mathbf{x}_1)V^\dag(\mathbf{x}_2)t^ae^{-i\mathbf{p}\cdot\mathbf{x}_1}e^{-i\mathbf{q}\cdot\mathbf{x}_2}  \nonumber \\ 
& \left[ \frac{ Q (z_1 )^{\frac32}\sqrt{z_2} (1-z_1)^2}{(2\pi)}\right] K_0\left(|\mathbf{x}_{12}|Q_1\right) \frac{e^{-i\mathbf{k}\cdot\mathbf{x}_2}}{(q+k)^2} \left( \frac{\mathbf{k}\cdot \boldsymbol{\epsilon}}{z_3} - \frac{ \mathbf{q}\cdot \boldsymbol{\epsilon}}{z_2}\right) \label{ia2final2}. \\
i\mathcal{M}_3^{a\, L;+,+} =& \frac{8egl^+}{z_1} \int \dd^2 \mathbf{x}_1 \dd^2\mathbf{x}_2 \dd^2 \mathbf{x}_3  V(\mathbf{x}_1)t^bV^\dag(\mathbf{x}_2)U(\mathbf{x}_3)_{ba}\,e^{-i\mathbf{p}\cdot\mathbf{x}_1}e^{-i\mathbf{q}\cdot\mathbf{x}_2}  \nonumber \\
& \left[- Q(z_1z_2)^{\frac32}(1-z_2)\right] \frac{ i}{(2\pi)^2} \frac{z_1}{1-z_2} \frac{ \mathbf{x}_{31}\cdot \boldsymbol{\epsilon}}{\mathbf{x}_{31}^2} K_0(QX) \, e^{-i\mathbf{k}\cdot\mathbf{x}_3} .\label{ia3final2}\\
i\mathcal{M}_4^{a\,L;+,+} =& \frac{8egl^+}{z_2} \int \dd^2\mathbf{x}_1 \dd^2\mathbf{x}_2\dd^2 \mathbf{x}_3 V(\mathbf{x}_1)t^bV^\dag(\mathbf{x}_2)U(\mathbf{x}_3)_{ba} \,e^{-i\mathbf{p}\cdot\mathbf{x}_1}e^{-i\mathbf{q}\cdot\mathbf{x}_2} \nonumber \\
&  \left[ Q(z_1)^{\frac32}\sqrt{z_2}(1-z_1)^2\right] \frac{ i}{(2\pi)^2} \frac{z_2}{1-z_1} \frac{ \mathbf{x}_{32}\cdot \boldsymbol{\epsilon}}{\mathbf{x}_{32}^2} K_0(QX) \, e^{-i\mathbf{k}\cdot\mathbf{x}_3} \label{ia4final2}
\end{align}
\noindent where we have defined some shorthand notations,

\begin{align}
X = \sqrt{z_2z_1 \mathbf{x}_{12}^2 + z_3 z_1 \mathbf{x}_{13}^2 +z_3z_2 \mathbf{x}_{23}^2}, \,\,\,\,\,\,\,\,\,\,\, \mathbf{x}_{ij} = \mathbf{x}_i - \mathbf{x}_j, \,\,\,\,\,\,\,\,\,\,\, Q_i = Q\sqrt{z_i(1-z_i)}, \,\,\,\,\,\,\,\,\,\,\,\, \boldsymbol{\epsilon} = \frac{1}{\sqrt{2}} (1,i).
\end{align}

The next step is to square a given helicity amplitude and then to add up 
all the squared helicity amplitudes to get the un-polarized cross section. 
For example, the first helicity amplitude (\ref{ia1final2}) squared and 
summed over final state helicities, labeled as $\mathcal{M}_{11}^{L}$ is  

\begin{align}
\mathcal{M}_{11}^{L} =& \frac{ 64e^2g^2 Q^2(l^+)^2 z_1z_2^3(1-z_2)^2(z_1^2+(1-z_2)^2)}{(2\pi)^2} \int \dd^8 x \Tr_c\left[ t^a(V_1V_2^\dag - 1)(V_2^\prime V_1^{\prime \dag} - 1)t^a\right] e^{i\mathbf{p}\cdot(\mathbf{x}_1^\prime-\mathbf{x}_1)}e^{i\mathbf{q}\cdot(\mathbf{x}_2^\prime-\mathbf{x}_2)}\nonumber \\ 
& K_0(|\mathbf{x}_{12}|Q_2) K_0(|\mathbf{x}_{1^\prime 2^\prime}| Q_2) \frac{ e^{i\mathbf{k}\cdot(\mathbf{x}_1^\prime-\mathbf{x}_1)}}{(p+k)^4} \left( \frac{ \mathbf{p}}{z_1} - \frac{\mathbf{k}}{z_3}\right)^2 \label{m11} 
\end{align}
where $\dd^8 x$ denotes $\dd^2 \mathbf{x}_1 \dd^2 \mathbf{x}_2 \dd^2 \mathbf{x}_1^\prime \dd^2 \mathbf{x}_2^\prime$ with $V_i \equiv V(\mathbf{x}_i)$. To proceed further we take the soft
gluon limit ($z_3 \ll z_1,z_2$) and perform the integrals over the transverse
momentum $\mathbf{k}$. We also define the two and four point functions of Wilson lines 
 $S^{(2)}(\mathbf{x}_i,\mathbf{x}_j)$ with shorthand $S_{ij}$ and  $S^{(4)}(\mathbf{x}_i,\mathbf{x}_j,\mathbf{x}_k,\mathbf{x}_l)$ with shorthand $S_{ijkl}$ (dipoles and 
quadrupoles) as 
\begin{align}
S_{ij} = \frac{1}{N_c} \Tr_c \left\langle V(\mathbf{x}_i)V^\dag(\mathbf{x}_j)\right\rangle, \,\,\,\,\,\,\,\,\, S_{ijkl} = \frac{1}{N_c} \Tr_c \left\langle V(\mathbf{x}_i)V^\dag(\mathbf{x}_j)V(\mathbf{x}_k)V^\dag(\mathbf{x}_l\right\rangle.
\end{align}
which can be evaluated explicitly in the Gaussian 
approximation~\cite{Blaizot:2004wv,JalilianMarian:2004da,Dominguez:2011wm,Fukushima:2017mko,Lappi:2020srm} 
and are given by 

\begin{align}
S_{ij} = e^{-Q_s^2 \Gamma_{ij}}, \,\,\,\,\,\,\, S_{ijkl} = S_{il}S_{jk} - \left[ \frac{ \Gamma_{ij}^2 + \Gamma_{kl}^2 - \Gamma_{ik}^2 - \Gamma_{jl}^2}{\Gamma_{ij}^2 + \Gamma_{kl}^2 - \Gamma_{il}^2 - \Gamma_{jk}^2}\right]\left( S_{il}S_{jk} - S_{ij}S_{kl}\right), \,\,\,\,\,\, \Gamma_{ij} = (\mathbf{x}_i - \mathbf{x}_j)^2 \log \frac{1}{\Lambda|\mathbf{x}_i - \mathbf{x}_j|}.
\end{align}
where $\Lambda$ is an infrared regulator. The presence of the logarithm in $\Gamma_{ij}$  is essential for the correct power-law behavior of the cross sections at high transverse momenta. However it does make the analytic evaluation of these 
integrals impossible. As we will be exploring the more interesting 
(and experimentally accessible) region of low to intermediate transverse momenta we
will ignore it in this exploratory study which corresponds to taking the
Golec Biernat-Wusthoff model~\cite{GolecBiernat:1998js} of the dipole 
profile. With these approximations our full amplitude squared for 
production of a quark, anti-quark and a gluon with the transverse 
momentum of the gluon integrated can be written as 

\begin{align}
\int \frac{\dd^2 \mathbf{k}}{(2\pi)^2} |i\mathcal{M}|^2 =\int \frac{\dd^2 \mathbf{k}}{(2\pi)^2} \bigg[ \mathcal{M}_{11}^L + &\mathcal{M}_{22}^L +\mathcal{M}_{33}^L + \mathcal{M}_{44}^L +2 \mathcal{M}_{12}^L +2\mathcal{M}_{13}^L + 2\mathcal{M}_{14}^L + 2\mathcal{M}_{23}^L + 2\mathcal{M}_{24}^L + 2\mathcal{M}_{34}^L \bigg]  \nonumber \\
=& \frac{64 e^2 g^2 Q^2(l^+)^2N_c^2 (z_1 z_2)^3}{(2\pi)^4} \int \dd^{10}x K_0(|\mathbf{x}_{12}|Q_1)K_0(|\mathbf{x}_{1^\prime 2^\prime}|Q_1) e^{i\mathbf{p}\cdot(\mathbf{x}_1^\prime - \mathbf{x}_1)}e^{i\mathbf{q}\cdot(\mathbf{x}_2^\prime - \mathbf{x}_2)} \nonumber \\
&\bigg\{\left[ S_{122^\prime 1^\prime} - S_{12} - S_{1^\prime 2^\prime} + 1\right]\left(\Delta^{(3)}_{11^\prime } + \Delta^{(3)}_{22^\prime }\right) \nonumber \\
&+  \left[ S_{11^\prime} S_{22^\prime} \nonumber - S_{23}S_{13} - S_{2^\prime 3} S_{1^\prime 3} + 1 \right] \left( \Delta^{(3)}_{11^\prime} + \Delta^{(3)}_{22^\prime} - 2\Delta^{(3)}_{12^\prime}\right)\nonumber \\
 & -  \left[ S_{12}S_{1^\prime 2^\prime} - S_{12} - S_{1^\prime 2^\prime} + 1\right]2\Delta^{(3)}_{12^\prime}\nonumber \\ 
& +\left[ S_{122^\prime 3} S_{1^\prime 3} - S_{2^\prime 3}S_{1^\prime 3} - S_{12} + 1\right] \left(2\Delta^{(3)}_{11^\prime} - 2\Delta^{(3)}_{12^\prime}\right) \nonumber \\
& + \left[ S_{1231^\prime} S_{2^\prime 3} - S_{2^\prime 3} S_{1^\prime 3} - S_{12} + 1\right]\left(2\Delta^{(3)}_{22^\prime } - 2\Delta^{(3)}_{21^\prime}\right) \bigg\}. \label{combined}
\end{align}
with the radiation kernel $\Delta^{(3)}_{ij}$ given by 
\begin{align}
\Delta^{(3)}_{ij} = \frac{(\mathbf{x}_3 - \mathbf{x}_i)\cdot(\mathbf{x}_3 - \mathbf{x}_j)}{(\mathbf{x}_3 - \mathbf{x}_i)^2(\mathbf{x}_3 - \mathbf{x}_j)^2}.
\end{align}
While this result looks very compact it is still not very amenable to 
phenomenological studies of importance to experiments. Therefore we  
consider the more interesting limit of back-to-back azimuthal angular 
correlations.

\section{The back-to-back limit of azimuthal angular correlations}
To make a quantitative estimate of the role of coherent energy loss
and gluon saturation we will focus on the back-to-back kinematics region
in dihadron production in DIS at small $x$. To this end we define 
the total and relative momenta 
\begin{align}
\mathbf{P} \equiv \mathbf{p} + \mathbf{q}, \,\,\,\,\,\,\,\,  \mathbf{K} \equiv z_2 \mathbf{p} - z_1 \mathbf{q}.
\end{align}
and take the "back-to-back correlation" limit~\cite{Dominguez:2010xd,Dominguez:2011wm} defined as
\begin{align}
|\mathbf{P}| \ll  |\mathbf{K}| \sim |\mathbf{p}| \sim  |\mathbf{q}|
\end{align} 
We define new coordinate-space variables, $\mathbf{u},\mathbf{v},\mathbf{u}^\prime,\mathbf{v}^\prime$, 

\begin{align}
\mathbf{u} = \mathbf{x}_1 - \mathbf{x}_2, \,\,\,\,\, \mathbf{v} = z_1 \mathbf{x}_1 + z_2 \mathbf{x}_2, \,\,\,\,\, \mathbf{u}^\prime = \mathbf{x}_1^\prime - \mathbf{x}_2^\prime, \,\,\,\,\,\, \mathbf{v}^\prime = z_1 \mathbf{x}_1^\prime + z_2 \mathbf{x}_2^\prime. \label{uvcoords}
\end{align}
in terms of which the back-to-back limit corresponds to taking $|\mathbf{u}| \sim |\mathbf{u}^\prime| \ll 1$. Taking this limit in (\ref{combined}) and expanding to lowest nonzero order in $\mathbf{u}$ and $\mathbf{u}^\prime$ we get

\begin{align}
\int \frac{\dd^2 \mathbf{k}}{(2\pi)^2} |i\mathcal{M}|^2 =&  \frac{64 e^2 g^2 Q^2(l^+)^2N_c^2 (z_1 z_2)^3}{(2\pi)^4} \pi R^2 \int  \dd^2 \mathbf{u} \dd^2 \mathbf{u}^\prime \dd^2 \mathbf{v} \dd^2 \mathbf{v}^\prime \, K_0(Q_1|\mathbf{u}|)K_0(Q_1|\mathbf{u}^\prime|) e^{i\mathbf{P}\cdot(\mathbf{v}^\prime - \mathbf{v})} e^{i\mathbf{K}\cdot(\mathbf{u}^\prime - \mathbf{u})} \nonumber \\ \Bigg[&\frac{ (1-e^{-2Q_s^2(\mathbf{v}-\mathbf{v}^\prime)^2}) \left[ \mathbf{v}^2 + \mathbf{v}^{\prime 2 }- (\mathbf{v}-\mathbf{v}^\prime)^2\right] \mathbf{u}\cdot \mathbf{u}^\prime}{(\mathbf{v}-\mathbf{v}^\prime)^2\mathbf{v}^2\mathbf{v}^{\prime2}}  \nonumber \\&  +\frac{ 4e^{-2Q_s^2 \mathbf{v}^{\prime 2}} \left( 1 - e^{-2Q_s^2(\mathbf{v}^2  - \mathbf{v}\cdot \mathbf{v}^\prime)}\right) (\mathbf{u}\cdot \mathbf{v}^\prime) \left[2(\mathbf{v}\cdot \mathbf{v}^\prime)(\mathbf{u}^\prime \cdot \mathbf{v}^\prime) - (\mathbf{u}^\prime \cdot \mathbf{v}) \mathbf{v}^{\prime 2}\right]}{\mathbf{v}^2 \mathbf{v}^{\prime 4} (\mathbf{v}^2 - \mathbf{v} \cdot \mathbf{v}^\prime)} \nonumber \\ &+\frac{\left(1+e^{-2Q_s^2(\mathbf{v}-\mathbf{v}^\prime)^2}-e^{-2Q_s^2\mathbf{v}^2}-e^{-2Q_s^2\mathbf{v}^{\prime 2}}\right) }{\mathbf{v}^4\mathbf{v}^{\prime 4}}\left[ 4(\mathbf{u}\cdot \mathbf{v})(\mathbf{u}^\prime \cdot \mathbf{v}^\prime) (\mathbf{v}\cdot \mathbf{v}^\prime) - 4(\mathbf{u}\cdot \mathbf{v}^\prime) (\mathbf{u}^\prime \cdot \mathbf{v}^\prime) \mathbf{v}^2 + (\mathbf{u}\cdot \mathbf{u}^\prime) \mathbf{v}^2 \mathbf{v}^{\prime 2}\right] \Bigg]
\end{align}
where $\pi R^2$ is the transverse area of the target arising from integration
over $\mathbf{x}_3$ (after a substitution). The integrals over $\mathbf{u}$ and $\mathbf{u}^\prime$ can now be evaluated explicitly and the remaining integrals over $\mathbf{v}$ and $\mathbf{v}^\prime$ integrals can be written in polar coordinates to get 

\begin{align}
\int \frac{\dd^2 \mathbf{k}}{(2\pi)^2} |i\mathcal{M}|^2=& \frac{64 e^2 g^2 Q^2(l^+)^2N_c^2 (z_1 z_2)^3}{(2\pi)^2} \frac{4\pi R^2 K^2}{(K^2+Q_1^2)^4} \int  \dd r \dd r^\prime \dd \phi \dd \phi^\prime \, e^{i\frac{P}{\Lambda}r^\prime \cos( \phi^\prime - \phi_P)} e^{-i\frac{P}{\Lambda}r\cos(\phi - \phi_P)}  \nonumber \\ &\Bigg[2\frac{ (1-e^{-2\frac{Q_s}{\Lambda}(r^2+r^{\prime 2} - 2rr^\prime \cos(\phi - \phi^\prime)})\cos(\phi - \phi^\prime) }{r^2+r^{\prime 2} - 2rr^\prime \cos(\phi - \phi^\prime)}  \nonumber \\
&  +\frac{ 4e^{-2\frac{Q_s}{\Lambda} r^{\prime 2}} \left( 1 - e^{-2\frac{Q_s}{\Lambda}(r^2  - rr^\prime \cos(\phi - \phi^\prime))}\right) \cos(\phi^\prime) \left[2\cos(\phi-\phi^\prime)\cos(\phi^\prime) - \cos(\phi)\right]}{r^2 - rr^\prime\cos(\phi-\phi^\prime))} \nonumber \\
 &+\frac{\left(1+e^{-2\frac{Q_s}{\Lambda}(r^2+r^{\prime 2} - 2rr^\prime\cos(\phi-\phi^\prime)}-e^{-2\frac{Q_s}{\Lambda}r^2}-e^{-2\frac{Q_s}{\Lambda}r^{\prime 2}}\right) }{rr^{\prime }}\left[4\cos(\phi)\cos(\phi^\prime)\cos(\phi-\phi^\prime) - 4\cos^2(\phi^\prime) + 1\right] \Bigg] \label{intf}
\end{align}

\noindent where we have defined new dimensionless variables $r$ and $r^\prime$ 

\begin{align}
r = \Lambda|\mathbf{v}|, \,\,\,\,\,\,\, r^\prime =\Lambda |\mathbf{v}|^\prime
\end{align}

\noindent and $\Lambda$ is a constant parameter with units of mass. All angles are measured relative to the $\mathbf{K}$ vector so that 
$\phi_P$ is the angle between $\mathbf{P}$ and $\mathbf{K}$, and ($\phi$, $\phi^\prime$) are the angles of ($\mathbf{v}$, $\mathbf{v}^\prime$) with respect to $\mathbf{K}$. 

We calculate the medium-induced, coherent energy loss as defined in (\ref{eloss}). The induced radiation spectrum can be written as,

\begin{align}
z_3 \frac{ \dd I_{\text{ind}}}{ \dd z_3} 
=  \frac{ 2 \alpha_s N_c}{(2\pi)^3} \left[ \frac{ \int \dd r\, \dd r^\prime \, \dd \phi \, \dd \phi^\prime \, f(r,r^\prime,\phi,\phi^\prime)}{\int \frac{ \dd r}{r} J_0\left( \frac{ P}{\Lambda} r\right) \left( 1- e^{-2\frac{Q_s}{\Lambda}r^2}\right)}\Bigg\rvert_{A} -  \frac{ \int \dd r\, \dd r^\prime \, \dd \phi \, \dd \phi^\prime \, f(r,r^\prime,\phi,\phi^\prime)}{\int \frac{ \dd r}{r} J_0\left( \frac{ P}{\Lambda} r\right) \left( 1- e^{-2\frac{Q_s}{\Lambda}r^2}\right)}\Bigg\rvert_{p}\right]. \label{zdz}
\end{align}

Here $f(r,r^\prime,\phi,\phi^\prime)$ is the integrand in Eq. \ref{intf}
and the Leading Order result~\cite{Gelis:2010nm} is used in the denominator.
The only difference between the first and second terms is the different saturation scale $Q_s$ of a nucleus and a proton. It should be noted that we have taken the same back-to-back limit in the denominators above which describe the Leading Order
quark anti-quark production cross sections. Furthermore, we use a 
cut off on the integration variables $r$ and $r^\prime$ in order
to impose color neutrality at scales comparable to confinement scale
$\Lambda_{QCD}$ (i.e. we choose $\Lambda = \Lambda_{QCD} = 200\text{MeV}$ and the $r,r^\prime$ integrals then go from $0$ to $1$). 
Finally, to get an estimate of the energy loss effects we consider
some specific values for the final state partons; we will 
consider the case when both partons have similar rapidities
so that $z_1 = 0.55, z_2 = 0.45$ and their transverse momenta are of
the order of the photon virtuality $Q$, specifically

\begin{align}
|\mathbf{p}| = Q  \,\,\,\,\,\, |\mathbf{q}| = Q + 0.1\,\text{GeV}, \,\,\,\,\,\,\,\, Q^2_{s,p} = 1\, \text{GeV}^2, \,\,\,\,\,\,\, Q_{s,A}^2 = A^{1/3} Q^2_{s,p}.
\end{align}

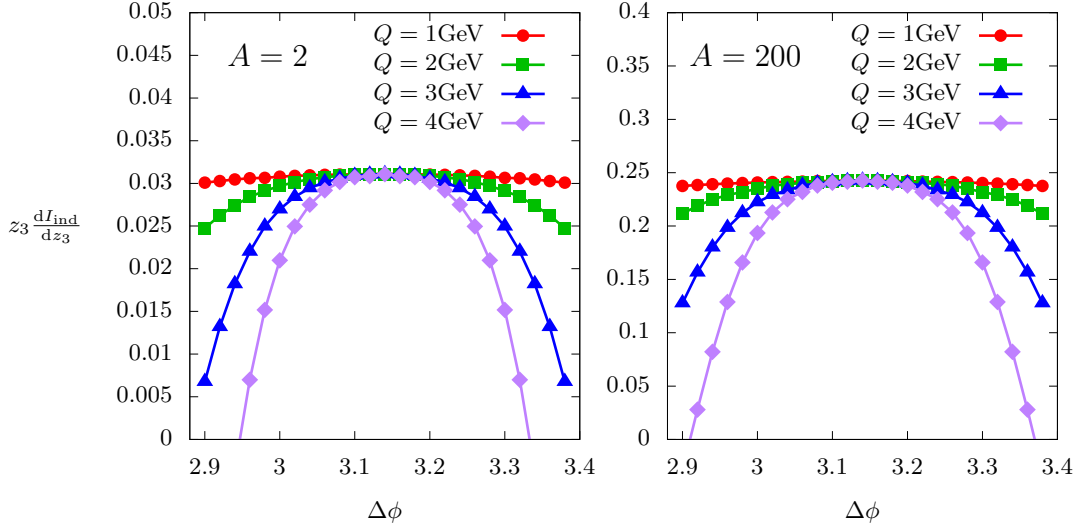
\begin{figure}
\begin{center}
\input{Elossplot}
\end{center}
\caption{The induced radiation spectrum plotted for $A=2$ and $A=200$ at four different values of $Q$. Note that we maintain $|\mathbf{p}| = Q$ and $|\mathbf{q}| = Q + 0.1\text{GeV}$ and that the vertical scales in the two graphs are different.} \label{elossplot}
\end{figure}

\noindent We show our results for the medium-induced coherent energy 
loss (\ref{zdz}) in Fig.(\ref{elossplot}) for various values of external momenta, and two different values of nuclear saturation
scale which mimics changing centrality, rapidity and/or nuclear $A$ number.

As seen the induced radiation is largest at the back to back kinematics and drops off sharply as one goes away from this limit. Also at the exact back to back limit the induced radiation is independent of photon virtuality, this is so since we have taken the quark and anti-quark momenta equal to photon virtuality. Furthermore the size of induced radiation increases with nuclear size as expected. This is also indicative of the size of the Next to Leading Order corrections, however keeping in 
mind that contributions of some of the Next to Leading terms cancel between 
a proton and a nucleus target in our definition of energy loss radiation 
spectrum in (\ref{eloss}). This clearly shows the importance of the 
full Next to Leading Order corrections to dihadron production cross 
section in DIS at small $x$. This is work in progress and will be reported elsewhere~\cite{nlo:fbjjm}. Another effect that is known to be important
is the Sudakov effect~\cite{Mueller:2012uf,Zheng:2014vka}, however this 
is beyond the scope of this work and will not be considered here.

To implement the cold matter energy loss effects in dihadron angular correlations, we calculate the coincidence probability (\ref{CP}) using numerical methods to evaluate the remaining integrals. Following~\cite{Albacete:2010pg}, we choose similar values for the external momentum windows, we choose $p_{\text{min}} = 2\,\text{GeV}, p_{\text{max}} = 10\,\text{GeV}, q_{\text{min}} = 1\,\text{GeV}, q_{\text{max}} = p.$ We also impose color neutrality at lengths beyond 1fm by using a cutoff on the $r, r^\prime$ integrals by choosing $\Lambda=\Lambda_{QCD} = 0.2\,\text{GeV}$ and setting $r_{\text{max}} = 1$. We fix the rapidities by choosing $z_1 = 0.55, z_2 = 0.45$. We use the back-to-back limit in both $N_{\text{pair}}$ and in the single inclusive $N_{\text{trig}}$~\footnote{We are grateful to C. Marquet for emphasizing this point to us.} .

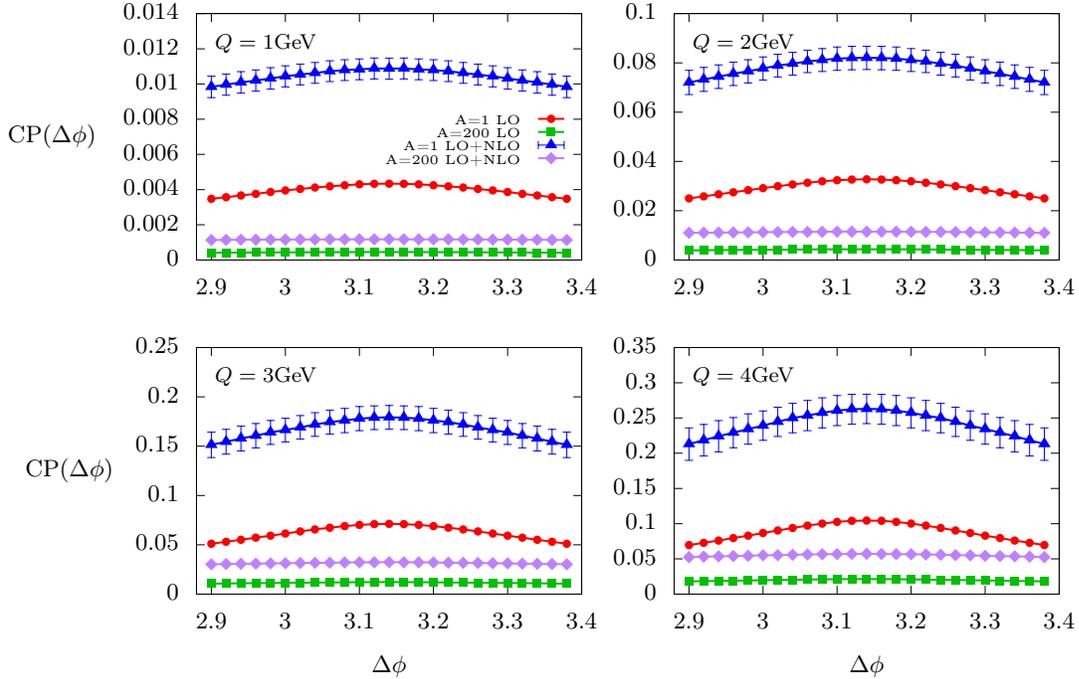
\begin{figure}[H]
\begin{center}
\input{CPplot}
\end{center}
\caption{The effect of Next to Leading order corrections on the coincidence probability CP$(\Delta\phi)$, shown for four different values of photon virtuality $Q$ for a proton target ($A=1$) and a large nuclear target ($A=200$).}\label{CPplot}
\end{figure}

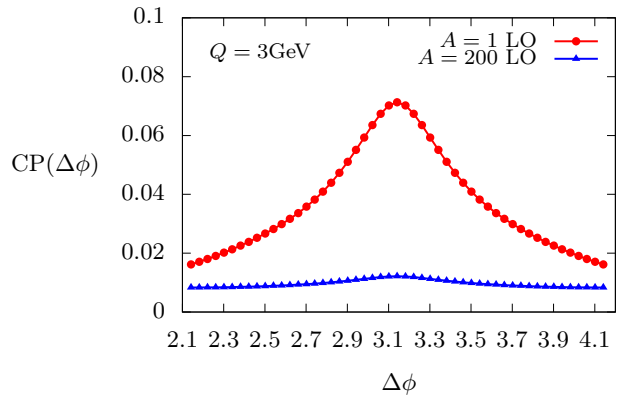
\begin{figure}[H]
\begin{center}
\input{CPplot4}
\end{center}
\caption{CP$(\Delta \phi)$ calculated at leading order using the back-to-back approximation in both $N_{\text{pair}}$ and $N_{\text{pair}}$ with a wider angle window $\Delta \phi \in [\pi - 1, \pi + 1]$ for the case $Q=3 \text{GeV}$.}\label{CPplot4}
\end{figure}

Clearly the induced radiation which is "lost" has a significant effect on the away side peak of dihadron azimuthal angular correlations. Note that we have restricted ourselves to angles in a very narrow window around the away side hadron. This is due to our strict back to back approximation which is expected to break down when going away from the away side hadron by a large angle. A proper quantitative estimate of the size of these corrections to the back to back approximation requires a detailed quantitative study using the various improved Transverse Momentum Dependent (TMD) distributions as advocated in~\cite{Altinoluk:2019wyu,Boussarie:2021lkb} and in~\cite{Kotko:2015ura,vanHameren:2016ftb,Altinoluk:2019fui,Boussarie:2020vzf,Fujii:2020bkl,Altinoluk:2021ygv} for proton-nucleus collisions. Nevertheless for the sake of comparison we show our results for a much wider range in $\Delta \phi$ once in Fig. (\ref{CPplot4}) and limit the rest of our analysis to the range $\Delta \phi \in [2.9,3.4]$ where we expect that the back-to-back approximation will still provide accurate results.

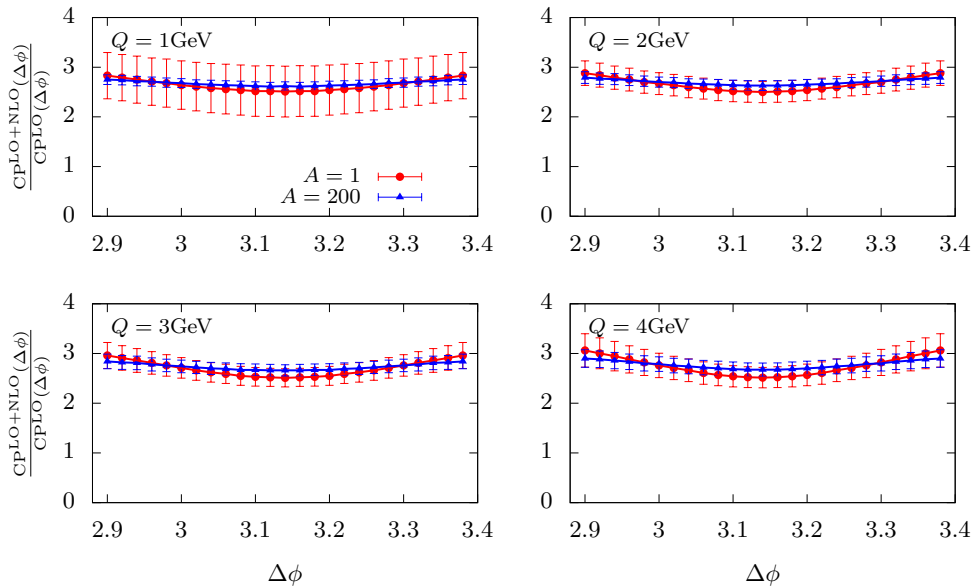
\begin{figure}[H]
\begin{center}
\input{CPplot3}
\end{center}
\caption{Here we show the ratio of CP$(\Delta\phi)$ at next-to-leading order versus at leading order calculated for a proton and a large nucleus target. A very weak dependence of next to leading order corrections on target size and angle is observed for small angles away from $\pi$.}\label{fig:CPplot3}
\end{figure}

To see the effect of the "lost" radiation more clearly we show the ratio of coincidence probabilities with the induced radiation to that of no radiation in Fig.(\ref{fig:CPplot3}) where an enhancement factor of order $3$ is seen with a weak angular dependence .

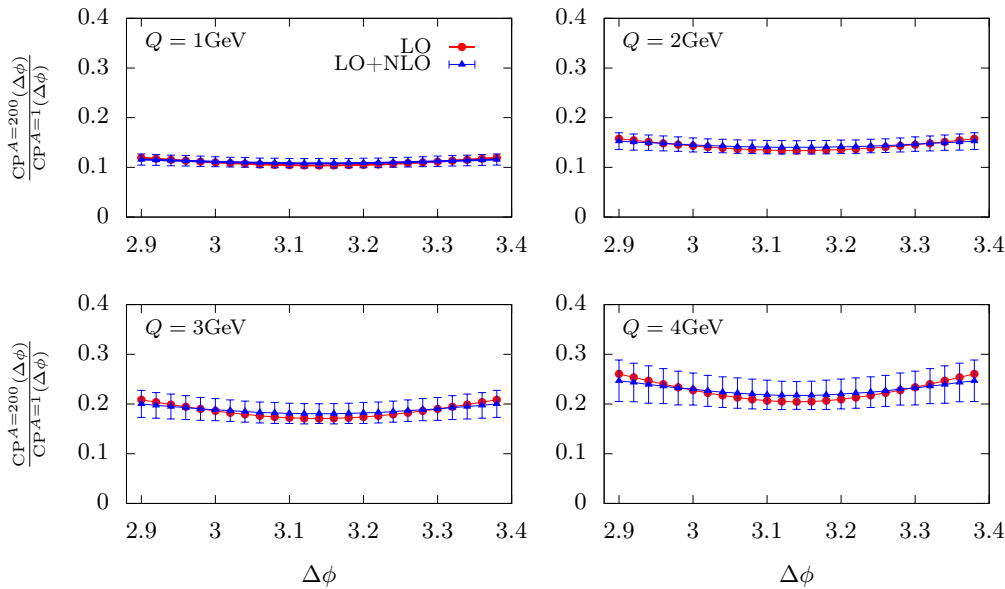
\begin{figure}[H]
\begin{center}
\input{CPplot2}
\end{center}
\caption{The double ratio of CP$(\Delta\phi)$ for a large nucleus over the CP$(\Delta\phi)$ for a proton, calculated at leading order and at next-to-leading order.}\label{fig:CPplot2}
\end{figure}

To investigate the medium dependence of the coincidence probability we define the double ratio of coincidence probabilities for a nucleus vs a proton in analogy with the medium modification factor $R_{pA}$ in proton-nucleus collisions and show this double ratio in Fig.(\ref{fig:CPplot2}). A significant reduction of the coincidence probability for a nucleus target is seen which is again very robust against next to leading order corrections. A clear increase in the magnitude of this double ratio is see with the increasing photon virtuality, reminiscent of the behavior of $R_{pA}$ in proton-nucleus collisions.

There are several ways in which our exploratory study can be improved; 
we have used the GBW model of dipole profile which is known to miss the 
high $p_t$ tail of the production spectra. One can improve the calculation
by using more realistic dipole profiles which have become available recently.
It will also be important to go beyond the strict back to back limit so that 
one can extend the present analysis to larger angles away from $\pi$. This 
will also shed light on the domain of applicability of back to back approximation.
Furthermore, we have only considered longitudinal photons, in a more realistic 
analysis one will need to include transverse photons as well. However we do not expect our results to change much by this. Lastly, we have considered the effect 
of radiated soft gluon on the quark anti-quark production by integrating out the radiated gluon. In a more realistic approach to dihadron production one will
need to consider integrating out any of the three outgoing partons. This will be
done when we compute the full next to leading order corrections to dihadron production and express (some of) the final state singularities into hadron fragmentation functions. This is work in progress and will be reported elsewhere~\cite{nlo:fbjjm}.

In summary we have performed an exploratory study of the 
contribution of coherent, medium-induced radiative energy loss 
to the away-side peak in dihadron azimuthal angular correlations 
in DIS at small $x$. We observe a sizable contribution to the 
coincidence probability from the induced radiation which indicates
the significance of next to leading order corrections to dihadron 
azimuthal angular correlations. We have defined 
a double ratio of coincidence probabilities and have shown that it is 
very stable against higher order corrections and thus may be a more
robust signature of saturation dynamics.

\vspace{-1ex}

\acknowledgments

We gratefully acknowledge support from the DOE Office of Nuclear Physics through Grant No. DE-SC0002307 and by PSC-CUNY through grant No. 63158-0051. We would like to thank F. Arleo, E. Aschenauer, G. Beuf, F. Gelis, K. Fukushima, T. Lappi, R. Venugopalan, B. Xiao and especially C. Marquet, S. Munier and S. Peign\'e for helpful discussions.

\bibliographystyle{apsrev}
\bibliography{01.Eloss}

\end{document}

%% file: Elossplot.tex
\begingroup
  \inputencoding{cp1252}%
  \makeatletter
  \providecommand\color[2][]{%
    \GenericError{(gnuplot) \space\space\space\@spaces}{%
      Package color not loaded in conjunction with
      terminal option `colourtext'%
    }{See the gnuplot documentation for explanation.%
    }{Either use 'blacktext' in gnuplot or load the package
      color.sty in LaTeX.}%
    \renewcommand\color[2][]{}%
  }%
  \providecommand\includegraphics[2][]{%
    \GenericError{(gnuplot) \space\space\space\@spaces}{%
      Package graphicx or graphics not loaded%
    }{See the gnuplot documentation for explanation.%
    }{The gnuplot epslatex terminal needs graphicx.sty or graphics.sty.}%
    \renewcommand\includegraphics[2][]{}%
  }%
  \providecommand\rotatebox[2]{#2}%
  \@ifundefined{ifGPcolor}{%
    \newif\ifGPcolor
    \GPcolorfalse
  }{}%
  \@ifundefined{ifGPblacktext}{%
    \newif\ifGPblacktext
    \GPblacktexttrue
  }{}%
  \let\gplgaddtomacro\g@addto@macro
  \gdef\gplbacktext{}%
  \gdef\gplfronttext{}%
  \makeatother
  \ifGPblacktext
    \def\colorrgb#1{}%
    \def\colorgray#1{}%
  \else
    \ifGPcolor
      \def\colorrgb#1{\color[rgb]{#1}}%
      \def\colorgray#1{\color[gray]{#1}}%
      \expandafter\def\csname LTw\endcsname{\color{white}}%
      \expandafter\def\csname LTb\endcsname{\color{black}}%
      \expandafter\def\csname LTa\endcsname{\color{black}}%
      \expandafter\def\csname LT0\endcsname{\color[rgb]{1,0,0}}%
      \expandafter\def\csname LT1\endcsname{\color[rgb]{0,1,0}}%
      \expandafter\def\csname LT2\endcsname{\color[rgb]{0,0,1}}%
      \expandafter\def\csname LT3\endcsname{\color[rgb]{1,0,1}}%
      \expandafter\def\csname LT4\endcsname{\color[rgb]{0,1,1}}%
      \expandafter\def\csname LT5\endcsname{\color[rgb]{1,1,0}}%
      \expandafter\def\csname LT6\endcsname{\color[rgb]{0,0,0}}%
      \expandafter\def\csname LT7\endcsname{\color[rgb]{1,0.3,0}}%
      \expandafter\def\csname LT8\endcsname{\color[rgb]{0.5,0.5,0.5}}%
    \else
      \def\colorrgb#1{\color{black}}%
      \def\colorgray#1{\color[gray]{#1}}%
      \expandafter\def\csname LTw\endcsname{\color{white}}%
      \expandafter\def\csname LTb\endcsname{\color{black}}%
      \expandafter\def\csname LTa\endcsname{\color{black}}%
      \expandafter\def\csname LT0\endcsname{\color{black}}%
      \expandafter\def\csname LT1\endcsname{\color{black}}%
      \expandafter\def\csname LT2\endcsname{\color{black}}%
      \expandafter\def\csname LT3\endcsname{\color{black}}%
      \expandafter\def\csname LT4\endcsname{\color{black}}%
      \expandafter\def\csname LT5\endcsname{\color{black}}%
      \expandafter\def\csname LT6\endcsname{\color{black}}%
      \expandafter\def\csname LT7\endcsname{\color{black}}%
      \expandafter\def\csname LT8\endcsname{\color{black}}%
    \fi
  \fi
    \setlength{\unitlength}{0.0500bp}%
    \ifx\gptboxheight\undefined%
      \newlength{\gptboxheight}%
      \newlength{\gptboxwidth}%
      \newsavebox{\gptboxtext}%
    \fi%
    \setlength{\fboxrule}{0.5pt}%
    \setlength{\fboxsep}{1pt}%
\begin{picture}(7200.00,4320.00)%
    \gplgaddtomacro\gplbacktext{%
      \csname LTb\endcsname
      \put(132,440){\makebox(0,0)[r]{\strut{}$0$}}%
      \put(132,762){\makebox(0,0)[r]{\strut{}$0.005$}}%
      \put(132,1084){\makebox(0,0)[r]{\strut{}$0.01$}}%
      \put(132,1406){\makebox(0,0)[r]{\strut{}$0.015$}}%
      \put(132,1728){\makebox(0,0)[r]{\strut{}$0.02$}}%
      \put(132,2050){\makebox(0,0)[r]{\strut{}$0.025$}}%
      \put(132,2371){\makebox(0,0)[r]{\strut{}$0.03$}}%
      \put(132,2693){\makebox(0,0)[r]{\strut{}$0.035$}}%
      \put(132,3015){\makebox(0,0)[r]{\strut{}$0.04$}}%
      \put(132,3337){\makebox(0,0)[r]{\strut{}$0.045$}}%
      \put(132,3659){\makebox(0,0)[r]{\strut{}$0.05$}}%
      \put(377,220){\makebox(0,0){\strut{}$2.9$}}%
      \put(942,220){\makebox(0,0){\strut{}$3$}}%
      \put(1507,220){\makebox(0,0){\strut{}$3.1$}}%
      \put(2073,220){\makebox(0,0){\strut{}$3.2$}}%
      \put(2638,220){\makebox(0,0){\strut{}$3.3$}}%
      \put(3203,220){\makebox(0,0){\strut{}$3.4$}}%
      \put(852,3337){\makebox(0,0){\large $A=2$}}%
    }%
    \gplgaddtomacro\gplfronttext{%
      \csname LTb\endcsname
      \put(-825,2049){\makebox(0,0){\strut{}$z_3 \frac{ \dd I_{\text{ind}}}{\dd z_3}$}}%
      \put(1733,-110){\makebox(0,0){\strut{}$\Delta \phi$}}%
      \csname LTb\endcsname
      \put(2480,3486){\makebox(0,0)[r]{\strut{}$Q=1$GeV}}%
      \csname LTb\endcsname
      \put(2480,3266){\makebox(0,0)[r]{\strut{}$Q=2$GeV}}%
      \csname LTb\endcsname
      \put(2480,3046){\makebox(0,0)[r]{\strut{}$Q=3$GeV}}%
      \csname LTb\endcsname
      \put(2480,2826){\makebox(0,0)[r]{\strut{}$Q=4$GeV}}%
    }%
    \gplgaddtomacro\gplbacktext{%
      \csname LTb\endcsname
      \put(3732,440){\makebox(0,0)[r]{\strut{}$0$}}%
      \put(3732,842){\makebox(0,0)[r]{\strut{}$0.05$}}%
      \put(3732,1245){\makebox(0,0)[r]{\strut{}$0.1$}}%
      \put(3732,1647){\makebox(0,0)[r]{\strut{}$0.15$}}%
      \put(3732,2050){\makebox(0,0)[r]{\strut{}$0.2$}}%
      \put(3732,2452){\makebox(0,0)[r]{\strut{}$0.25$}}%
      \put(3732,2854){\makebox(0,0)[r]{\strut{}$0.3$}}%
      \put(3732,3257){\makebox(0,0)[r]{\strut{}$0.35$}}%
      \put(3732,3659){\makebox(0,0)[r]{\strut{}$0.4$}}%
      \put(3977,220){\makebox(0,0){\strut{}$2.9$}}%
      \put(4542,220){\makebox(0,0){\strut{}$3$}}%
      \put(5107,220){\makebox(0,0){\strut{}$3.1$}}%
      \put(5673,220){\makebox(0,0){\strut{}$3.2$}}%
      \put(6238,220){\makebox(0,0){\strut{}$3.3$}}%
      \put(6803,220){\makebox(0,0){\strut{}$3.4$}}%
      \put(4452,3337){\makebox(0,0){\large $A=200$}}%
    }%
    \gplgaddtomacro\gplfronttext{%
      \csname LTb\endcsname
      \put(5333,-110){\makebox(0,0){\strut{}$\Delta \phi$}}%
      \csname LTb\endcsname
      \put(6080,3486){\makebox(0,0)[r]{\strut{}$Q=1$GeV}}%
      \csname LTb\endcsname
      \put(6080,3266){\makebox(0,0)[r]{\strut{}$Q=2$GeV}}%
      \csname LTb\endcsname
      \put(6080,3046){\makebox(0,0)[r]{\strut{}$Q=3$GeV}}%
      \csname LTb\endcsname
      \put(6080,2826){\makebox(0,0)[r]{\strut{}$Q=4$GeV}}%
    }%
    \gplbacktext
    \put(0,0){\includegraphics[width={360.00bp},height={216.00bp}]{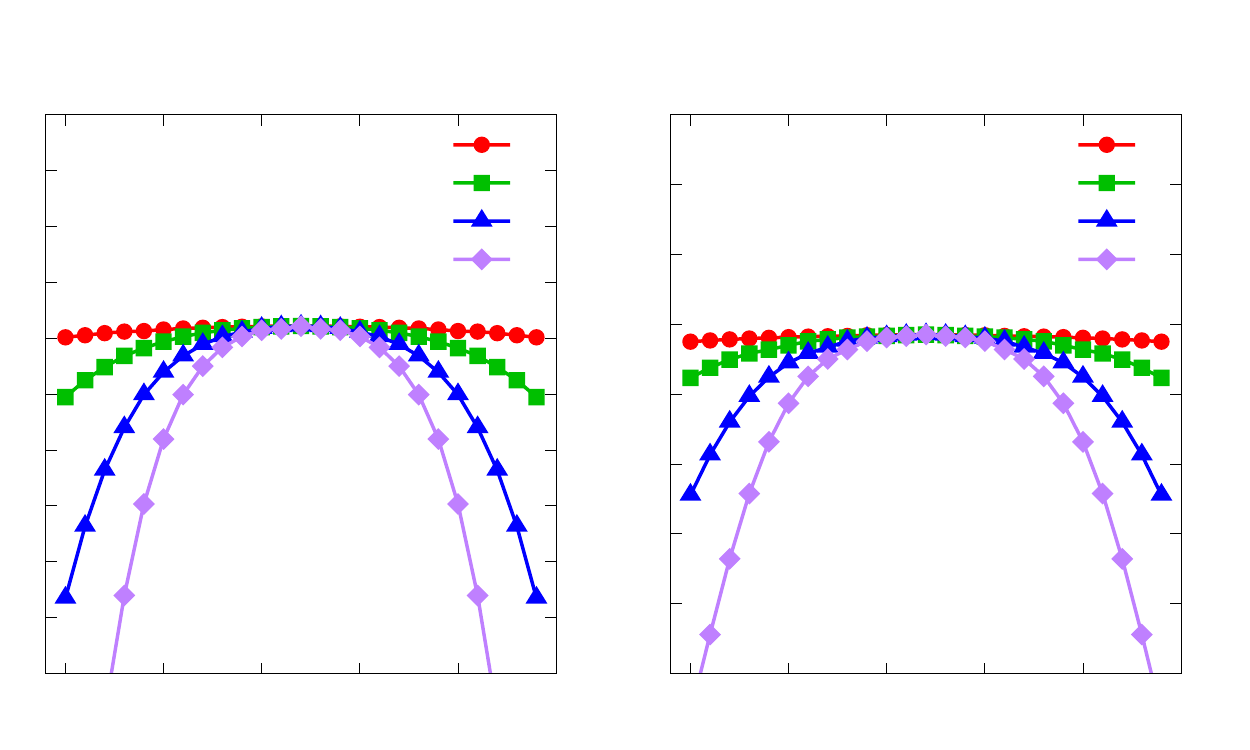}}%
    \gplfronttext
  \end{picture}%
\endgroup

%% file: CPplot.tex
\begingroup
  \inputencoding{cp1252}%
  \makeatletter
  \providecommand\color[2][]{%
    \GenericError{(gnuplot) \space\space\space\@spaces}{%
      Package color not loaded in conjunction with
      terminal option `colourtext'%
    }{See the gnuplot documentation for explanation.%
    }{Either use 'blacktext' in gnuplot or load the package
      color.sty in LaTeX.}%
    \renewcommand\color[2][]{}%
  }%
  \providecommand\includegraphics[2][]{%
    \GenericError{(gnuplot) \space\space\space\@spaces}{%
      Package graphicx or graphics not loaded%
    }{See the gnuplot documentation for explanation.%
    }{The gnuplot epslatex terminal needs graphicx.sty or graphics.sty.}%
    \renewcommand\includegraphics[2][]{}%
  }%
  \providecommand\rotatebox[2]{#2}%
  \@ifundefined{ifGPcolor}{%
    \newif\ifGPcolor
    \GPcolorfalse
  }{}%
  \@ifundefined{ifGPblacktext}{%
    \newif\ifGPblacktext
    \GPblacktexttrue
  }{}%
  \let\gplgaddtomacro\g@addto@macro
  \gdef\gplbacktext{}%
  \gdef\gplfronttext{}%
  \makeatother
  \ifGPblacktext
    \def\colorrgb#1{}%
    \def\colorgray#1{}%
  \else
    \ifGPcolor
      \def\colorrgb#1{\color[rgb]{#1}}%
      \def\colorgray#1{\color[gray]{#1}}%
      \expandafter\def\csname LTw\endcsname{\color{white}}%
      \expandafter\def\csname LTb\endcsname{\color{black}}%
      \expandafter\def\csname LTa\endcsname{\color{black}}%
      \expandafter\def\csname LT0\endcsname{\color[rgb]{1,0,0}}%
      \expandafter\def\csname LT1\endcsname{\color[rgb]{0,1,0}}%
      \expandafter\def\csname LT2\endcsname{\color[rgb]{0,0,1}}%
      \expandafter\def\csname LT3\endcsname{\color[rgb]{1,0,1}}%
      \expandafter\def\csname LT4\endcsname{\color[rgb]{0,1,1}}%
      \expandafter\def\csname LT5\endcsname{\color[rgb]{1,1,0}}%
      \expandafter\def\csname LT6\endcsname{\color[rgb]{0,0,0}}%
      \expandafter\def\csname LT7\endcsname{\color[rgb]{1,0.3,0}}%
      \expandafter\def\csname LT8\endcsname{\color[rgb]{0.5,0.5,0.5}}%
    \else
      \def\colorrgb#1{\color{black}}%
      \def\colorgray#1{\color[gray]{#1}}%
      \expandafter\def\csname LTw\endcsname{\color{white}}%
      \expandafter\def\csname LTb\endcsname{\color{black}}%
      \expandafter\def\csname LTa\endcsname{\color{black}}%
      \expandafter\def\csname LT0\endcsname{\color{black}}%
      \expandafter\def\csname LT1\endcsname{\color{black}}%
      \expandafter\def\csname LT2\endcsname{\color{black}}%
      \expandafter\def\csname LT3\endcsname{\color{black}}%
      \expandafter\def\csname LT4\endcsname{\color{black}}%
      \expandafter\def\csname LT5\endcsname{\color{black}}%
      \expandafter\def\csname LT6\endcsname{\color{black}}%
      \expandafter\def\csname LT7\endcsname{\color{black}}%
      \expandafter\def\csname LT8\endcsname{\color{black}}%
    \fi
  \fi
    \setlength{\unitlength}{0.0500bp}%
    \ifx\gptboxheight\undefined%
      \newlength{\gptboxheight}%
      \newlength{\gptboxwidth}%
      \newsavebox{\gptboxtext}%
    \fi%
    \setlength{\fboxrule}{0.5pt}%
    \setlength{\fboxsep}{1pt}%
\begin{picture}(7200.00,5040.00)%
    \gplgaddtomacro\gplbacktext{%
      \csname LTb\endcsname
      \put(348,2960){\makebox(0,0)[r]{\strut{}$0$}}%
      \put(348,3226){\makebox(0,0)[r]{\strut{}$0.002$}}%
      \put(348,3491){\makebox(0,0)[r]{\strut{}$0.004$}}%
      \put(348,3757){\makebox(0,0)[r]{\strut{}$0.006$}}%
      \put(348,4022){\makebox(0,0)[r]{\strut{}$0.008$}}%
      \put(348,4288){\makebox(0,0)[r]{\strut{}$0.01$}}%
      \put(348,4553){\makebox(0,0)[r]{\strut{}$0.012$}}%
      \put(348,4819){\makebox(0,0)[r]{\strut{}$0.014$}}%
      \put(592,2740){\makebox(0,0){\strut{}$2.9$}}%
      \put(1150,2740){\makebox(0,0){\strut{}$3$}}%
      \put(1708,2740){\makebox(0,0){\strut{}$3.1$}}%
      \put(2266,2740){\makebox(0,0){\strut{}$3.2$}}%
      \put(2825,2740){\makebox(0,0){\strut{}$3.3$}}%
      \put(3383,2740){\makebox(0,0){\strut{}$3.4$}}%
      \put(1003,4596){\makebox(0,0){\footnotesize $Q=1\text{GeV}$}}%
    }%
    \gplgaddtomacro\gplfronttext{%
      \csname LTb\endcsname
      \put(-609,3889){\makebox(0,0){\strut{}CP$(\Delta \phi)$}}%
      \csname LTb\endcsname
      \put(2924,4025){\makebox(0,0)[r]{\strut{}\tiny{A=1 LO}}}%
      \csname LTb\endcsname
      \put(2924,3926){\makebox(0,0)[r]{\strut{}\tiny{A=200 LO}}}%
      \csname LTb\endcsname
      \put(2924,3827){\makebox(0,0)[r]{\strut{}\tiny{A=1 LO+NLO}}}%
      \csname LTb\endcsname
      \put(2924,3728){\makebox(0,0)[r]{\strut{}\tiny{A=200 LO+NLO}}}%
    }%
    \gplgaddtomacro\gplbacktext{%
      \csname LTb\endcsname
      \put(3948,2960){\makebox(0,0)[r]{\strut{}$0$}}%
      \put(3948,3332){\makebox(0,0)[r]{\strut{}$0.02$}}%
      \put(3948,3704){\makebox(0,0)[r]{\strut{}$0.04$}}%
      \put(3948,4075){\makebox(0,0)[r]{\strut{}$0.06$}}%
      \put(3948,4447){\makebox(0,0)[r]{\strut{}$0.08$}}%
      \put(3948,4819){\makebox(0,0)[r]{\strut{}$0.1$}}%
      \put(4192,2740){\makebox(0,0){\strut{}$2.9$}}%
      \put(4750,2740){\makebox(0,0){\strut{}$3$}}%
      \put(5308,2740){\makebox(0,0){\strut{}$3.1$}}%
      \put(5866,2740){\makebox(0,0){\strut{}$3.2$}}%
      \put(6424,2740){\makebox(0,0){\strut{}$3.3$}}%
      \put(6982,2740){\makebox(0,0){\strut{}$3.4$}}%
      \put(4602,4596){\makebox(0,0){\footnotesize $Q=2\text{GeV}$}}%
    }%
    \gplgaddtomacro\gplfronttext{%
    }%
    \gplgaddtomacro\gplbacktext{%
      \csname LTb\endcsname
      \put(348,440){\makebox(0,0)[r]{\strut{}$0$}}%
      \put(348,812){\makebox(0,0)[r]{\strut{}$0.05$}}%
      \put(348,1184){\makebox(0,0)[r]{\strut{}$0.1$}}%
      \put(348,1556){\makebox(0,0)[r]{\strut{}$0.15$}}%
      \put(348,1928){\makebox(0,0)[r]{\strut{}$0.2$}}%
      \put(348,2300){\makebox(0,0)[r]{\strut{}$0.25$}}%
      \put(592,220){\makebox(0,0){\strut{}$2.9$}}%
      \put(1150,220){\makebox(0,0){\strut{}$3$}}%
      \put(1708,220){\makebox(0,0){\strut{}$3.1$}}%
      \put(2266,220){\makebox(0,0){\strut{}$3.2$}}%
      \put(2825,220){\makebox(0,0){\strut{}$3.3$}}%
      \put(3383,220){\makebox(0,0){\strut{}$3.4$}}%
      \put(1003,2077){\makebox(0,0){\footnotesize $Q=3\text{GeV}$}}%
    }%
    \gplgaddtomacro\gplfronttext{%
      \csname LTb\endcsname
      \put(-477,1370){\makebox(0,0){\strut{}CP$(\Delta \phi)$}}%
      \put(1931,-110){\makebox(0,0){\strut{}$\Delta \phi$}}%
    }%
    \gplgaddtomacro\gplbacktext{%
      \csname LTb\endcsname
      \put(3948,440){\makebox(0,0)[r]{\strut{}$0$}}%
      \put(3948,706){\makebox(0,0)[r]{\strut{}$0.05$}}%
      \put(3948,971){\makebox(0,0)[r]{\strut{}$0.1$}}%
      \put(3948,1237){\makebox(0,0)[r]{\strut{}$0.15$}}%
      \put(3948,1503){\makebox(0,0)[r]{\strut{}$0.2$}}%
      \put(3948,1769){\makebox(0,0)[r]{\strut{}$0.25$}}%
      \put(3948,2034){\makebox(0,0)[r]{\strut{}$0.3$}}%
      \put(3948,2300){\makebox(0,0)[r]{\strut{}$0.35$}}%
      \put(4192,220){\makebox(0,0){\strut{}$2.9$}}%
      \put(4750,220){\makebox(0,0){\strut{}$3$}}%
      \put(5308,220){\makebox(0,0){\strut{}$3.1$}}%
      \put(5866,220){\makebox(0,0){\strut{}$3.2$}}%
      \put(6424,220){\makebox(0,0){\strut{}$3.3$}}%
      \put(6982,220){\makebox(0,0){\strut{}$3.4$}}%
      \put(4602,2077){\makebox(0,0){\footnotesize $Q=4\text{GeV}$}}%
    }%
    \gplgaddtomacro\gplfronttext{%
      \csname LTb\endcsname
      \put(5531,-110){\makebox(0,0){\strut{}$\Delta \phi$}}%
    }%
    \gplbacktext
    \put(0,0){\includegraphics[width={360.00bp},height={252.00bp}]{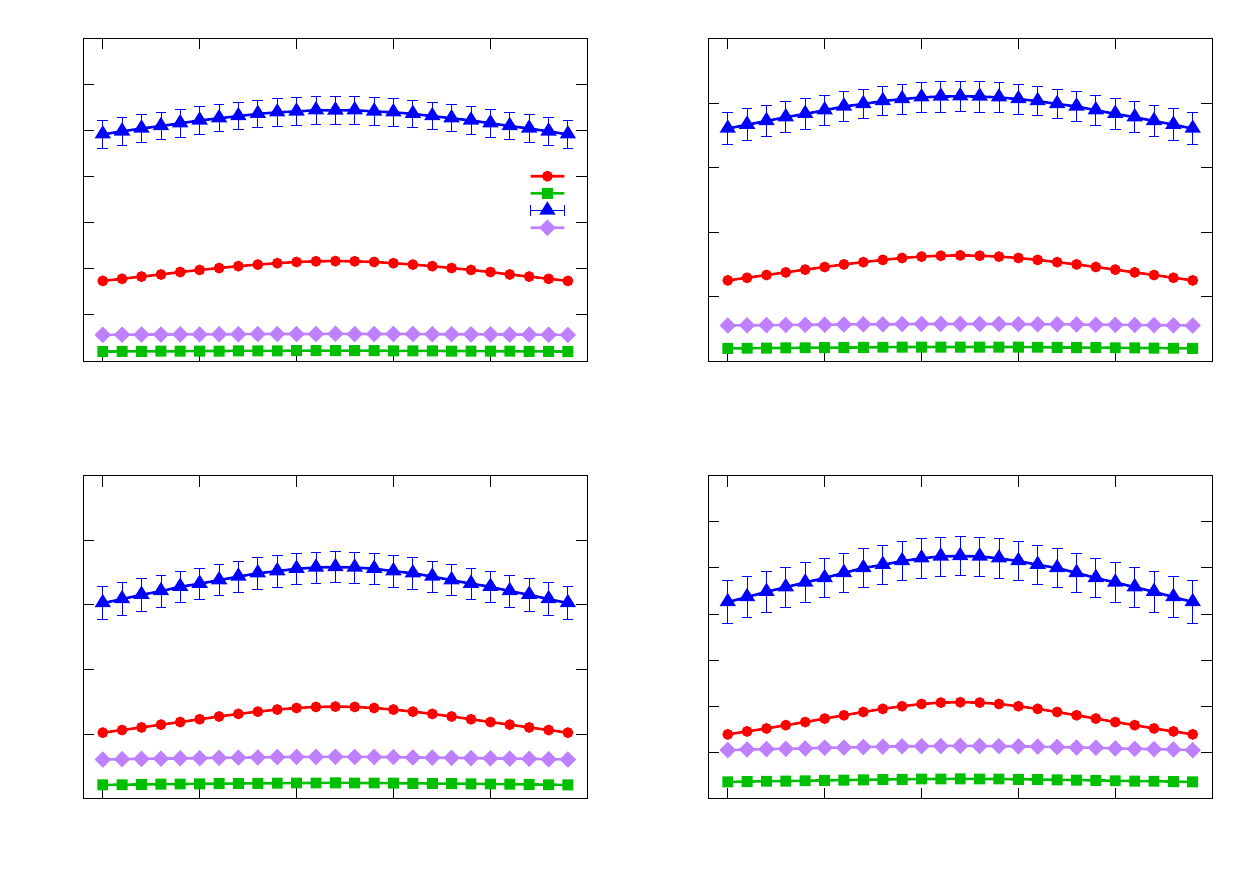}}%
    \gplfronttext
  \end{picture}%
\endgroup

%% file: CPplot4.tex
\begingroup
  \inputencoding{cp1252}%
  \makeatletter
  \providecommand\color[2][]{%
    \GenericError{(gnuplot) \space\space\space\@spaces}{%
      Package color not loaded in conjunction with
      terminal option `colourtext'%
    }{See the gnuplot documentation for explanation.%
    }{Either use 'blacktext' in gnuplot or load the package
      color.sty in LaTeX.}%
    \renewcommand\color[2][]{}%
  }%
  \providecommand\includegraphics[2][]{%
    \GenericError{(gnuplot) \space\space\space\@spaces}{%
      Package graphicx or graphics not loaded%
    }{See the gnuplot documentation for explanation.%
    }{The gnuplot epslatex terminal needs graphicx.sty or graphics.sty.}%
    \renewcommand\includegraphics[2][]{}%
  }%
  \providecommand\rotatebox[2]{#2}%
  \@ifundefined{ifGPcolor}{%
    \newif\ifGPcolor
    \GPcolorfalse
  }{}%
  \@ifundefined{ifGPblacktext}{%
    \newif\ifGPblacktext
    \GPblacktexttrue
  }{}%
  \let\gplgaddtomacro\g@addto@macro
  \gdef\gplbacktext{}%
  \gdef\gplfronttext{}%
  \makeatother
  \ifGPblacktext
    \def\colorrgb#1{}%
    \def\colorgray#1{}%
  \else
    \ifGPcolor
      \def\colorrgb#1{\color[rgb]{#1}}%
      \def\colorgray#1{\color[gray]{#1}}%
      \expandafter\def\csname LTw\endcsname{\color{white}}%
      \expandafter\def\csname LTb\endcsname{\color{black}}%
      \expandafter\def\csname LTa\endcsname{\color{black}}%
      \expandafter\def\csname LT0\endcsname{\color[rgb]{1,0,0}}%
      \expandafter\def\csname LT1\endcsname{\color[rgb]{0,1,0}}%
      \expandafter\def\csname LT2\endcsname{\color[rgb]{0,0,1}}%
      \expandafter\def\csname LT3\endcsname{\color[rgb]{1,0,1}}%
      \expandafter\def\csname LT4\endcsname{\color[rgb]{0,1,1}}%
      \expandafter\def\csname LT5\endcsname{\color[rgb]{1,1,0}}%
      \expandafter\def\csname LT6\endcsname{\color[rgb]{0,0,0}}%
      \expandafter\def\csname LT7\endcsname{\color[rgb]{1,0.3,0}}%
      \expandafter\def\csname LT8\endcsname{\color[rgb]{0.5,0.5,0.5}}%
    \else
      \def\colorrgb#1{\color{black}}%
      \def\colorgray#1{\color[gray]{#1}}%
      \expandafter\def\csname LTw\endcsname{\color{white}}%
      \expandafter\def\csname LTb\endcsname{\color{black}}%
      \expandafter\def\csname LTa\endcsname{\color{black}}%
      \expandafter\def\csname LT0\endcsname{\color{black}}%
      \expandafter\def\csname LT1\endcsname{\color{black}}%
      \expandafter\def\csname LT2\endcsname{\color{black}}%
      \expandafter\def\csname LT3\endcsname{\color{black}}%
      \expandafter\def\csname LT4\endcsname{\color{black}}%
      \expandafter\def\csname LT5\endcsname{\color{black}}%
      \expandafter\def\csname LT6\endcsname{\color{black}}%
      \expandafter\def\csname LT7\endcsname{\color{black}}%
      \expandafter\def\csname LT8\endcsname{\color{black}}%
    \fi
  \fi
    \setlength{\unitlength}{0.0500bp}%
    \ifx\gptboxheight\undefined%
      \newlength{\gptboxheight}%
      \newlength{\gptboxwidth}%
      \newsavebox{\gptboxtext}%
    \fi%
    \setlength{\fboxrule}{0.5pt}%
    \setlength{\fboxsep}{1pt}%
\begin{picture}(4320.00,2880.00)%
    \gplgaddtomacro\gplbacktext{%
      \csname LTb\endcsname
      \put(528,440){\makebox(0,0)[r]{\strut{}$0$}}%
      \put(528,884){\makebox(0,0)[r]{\strut{}$0.02$}}%
      \put(528,1328){\makebox(0,0)[r]{\strut{}$0.04$}}%
      \put(528,1771){\makebox(0,0)[r]{\strut{}$0.06$}}%
      \put(528,2215){\makebox(0,0)[r]{\strut{}$0.08$}}%
      \put(528,2659){\makebox(0,0)[r]{\strut{}$0.1$}}%
      \put(660,220){\makebox(0,0){\strut{}$2.1$}}%
      \put(971,220){\makebox(0,0){\strut{}$2.3$}}%
      \put(1282,220){\makebox(0,0){\strut{}$2.5$}}%
      \put(1592,220){\makebox(0,0){\strut{}$2.7$}}%
      \put(1903,220){\makebox(0,0){\strut{}$2.9$}}%
      \put(2214,220){\makebox(0,0){\strut{}$3.1$}}%
      \put(2525,220){\makebox(0,0){\strut{}$3.3$}}%
      \put(2835,220){\makebox(0,0){\strut{}$3.5$}}%
      \put(3146,220){\makebox(0,0){\strut{}$3.7$}}%
      \put(3457,220){\makebox(0,0){\strut{}$3.9$}}%
      \put(3768,220){\makebox(0,0){\strut{}$4.1$}}%
      \put(1247,2393){\makebox(0,0){\footnotesize $Q=3\text{GeV}$}}%
    }%
    \gplgaddtomacro\gplfronttext{%
      \csname LTb\endcsname
      \put(-297,1549){\makebox(0,0){\strut{}CP$(\Delta \phi)$}}%
      \put(2291,-110){\makebox(0,0){\strut{}$\Delta \phi$}}%
      \csname LTb\endcsname
      \put(3332,2482){\makebox(0,0)[r]{\strut{}\footnotesize{$A=1$ LO}}}%
      \csname LTb\endcsname
      \put(3332,2350){\makebox(0,0)[r]{\strut{}\footnotesize{$A=200$ LO}}}%
    }%
    \gplbacktext
    \put(0,0){\includegraphics[width={216.00bp},height={144.00bp}]{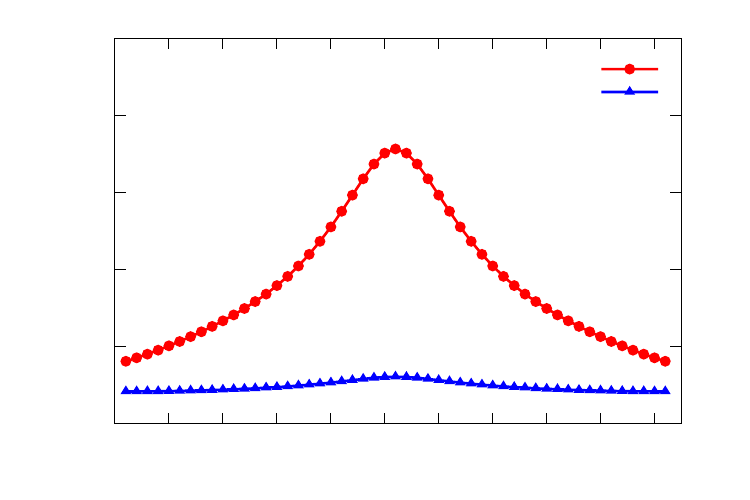}}%
    \gplfronttext
  \end{picture}%
\endgroup

%% file: CPplot3.tex
\begingroup
  \inputencoding{cp1252}%
  \makeatletter
  \providecommand\color[2][]{%
    \GenericError{(gnuplot) \space\space\space\@spaces}{%
      Package color not loaded in conjunction with
      terminal option `colourtext'%
    }{See the gnuplot documentation for explanation.%
    }{Either use 'blacktext' in gnuplot or load the package
      color.sty in LaTeX.}%
    \renewcommand\color[2][]{}%
  }%
  \providecommand\includegraphics[2][]{%
    \GenericError{(gnuplot) \space\space\space\@spaces}{%
      Package graphicx or graphics not loaded%
    }{See the gnuplot documentation for explanation.%
    }{The gnuplot epslatex terminal needs graphicx.sty or graphics.sty.}%
    \renewcommand\includegraphics[2][]{}%
  }%
  \providecommand\rotatebox[2]{#2}%
  \@ifundefined{ifGPcolor}{%
    \newif\ifGPcolor
    \GPcolorfalse
  }{}%
  \@ifundefined{ifGPblacktext}{%
    \newif\ifGPblacktext
    \GPblacktexttrue
  }{}%
  \let\gplgaddtomacro\g@addto@macro
  \gdef\gplbacktext{}%
  \gdef\gplfronttext{}%
  \makeatother
  \ifGPblacktext
    \def\colorrgb#1{}%
    \def\colorgray#1{}%
  \else
    \ifGPcolor
      \def\colorrgb#1{\color[rgb]{#1}}%
      \def\colorgray#1{\color[gray]{#1}}%
      \expandafter\def\csname LTw\endcsname{\color{white}}%
      \expandafter\def\csname LTb\endcsname{\color{black}}%
      \expandafter\def\csname LTa\endcsname{\color{black}}%
      \expandafter\def\csname LT0\endcsname{\color[rgb]{1,0,0}}%
      \expandafter\def\csname LT1\endcsname{\color[rgb]{0,1,0}}%
      \expandafter\def\csname LT2\endcsname{\color[rgb]{0,0,1}}%
      \expandafter\def\csname LT3\endcsname{\color[rgb]{1,0,1}}%
      \expandafter\def\csname LT4\endcsname{\color[rgb]{0,1,1}}%
      \expandafter\def\csname LT5\endcsname{\color[rgb]{1,1,0}}%
      \expandafter\def\csname LT6\endcsname{\color[rgb]{0,0,0}}%
      \expandafter\def\csname LT7\endcsname{\color[rgb]{1,0.3,0}}%
      \expandafter\def\csname LT8\endcsname{\color[rgb]{0.5,0.5,0.5}}%
    \else
      \def\colorrgb#1{\color{black}}%
      \def\colorgray#1{\color[gray]{#1}}%
      \expandafter\def\csname LTw\endcsname{\color{white}}%
      \expandafter\def\csname LTb\endcsname{\color{black}}%
      \expandafter\def\csname LTa\endcsname{\color{black}}%
      \expandafter\def\csname LT0\endcsname{\color{black}}%
      \expandafter\def\csname LT1\endcsname{\color{black}}%
      \expandafter\def\csname LT2\endcsname{\color{black}}%
      \expandafter\def\csname LT3\endcsname{\color{black}}%
      \expandafter\def\csname LT4\endcsname{\color{black}}%
      \expandafter\def\csname LT5\endcsname{\color{black}}%
      \expandafter\def\csname LT6\endcsname{\color{black}}%
      \expandafter\def\csname LT7\endcsname{\color{black}}%
      \expandafter\def\csname LT8\endcsname{\color{black}}%
    \fi
  \fi
    \setlength{\unitlength}{0.0500bp}%
    \ifx\gptboxheight\undefined%
      \newlength{\gptboxheight}%
      \newlength{\gptboxwidth}%
      \newsavebox{\gptboxtext}%
    \fi%
    \setlength{\fboxrule}{0.5pt}%
    \setlength{\fboxsep}{1pt}%
\begin{picture}(7200.00,4320.00)%
    \gplgaddtomacro\gplbacktext{%
      \csname LTb\endcsname
      \put(348,2600){\makebox(0,0)[r]{\strut{}$0$}}%
      \put(348,2975){\makebox(0,0)[r]{\strut{}$1$}}%
      \put(348,3350){\makebox(0,0)[r]{\strut{}$2$}}%
      \put(348,3724){\makebox(0,0)[r]{\strut{}$3$}}%
      \put(348,4099){\makebox(0,0)[r]{\strut{}$4$}}%
      \put(592,2380){\makebox(0,0){\strut{}$2.9$}}%
      \put(1150,2380){\makebox(0,0){\strut{}$3$}}%
      \put(1708,2380){\makebox(0,0){\strut{}$3.1$}}%
      \put(2266,2380){\makebox(0,0){\strut{}$3.2$}}%
      \put(2825,2380){\makebox(0,0){\strut{}$3.3$}}%
      \put(3383,2380){\makebox(0,0){\strut{}$3.4$}}%
      \put(1003,3919){\makebox(0,0){\footnotesize $Q=1$GeV}}%
    }%
    \gplgaddtomacro\gplfronttext{%
      \csname LTb\endcsname
      \put(7,3349){\rotatebox{-270}{\makebox(0,0){\strut{}$\frac{\text{CP}^{\text{LO}+\text{NLO}}(\Delta \phi)}{\text{CP}^{\text{LO}}(\Delta\phi)}$}}}%
      \csname LTb\endcsname
      \put(2502,2898){\makebox(0,0)[r]{\strut{}\footnotesize{$A=1$}}}%
      \csname LTb\endcsname
      \put(2502,2744){\makebox(0,0)[r]{\strut{}\footnotesize{$A=200$}}}%
    }%
    \gplgaddtomacro\gplbacktext{%
      \csname LTb\endcsname
      \put(3948,2600){\makebox(0,0)[r]{\strut{}$0$}}%
      \put(3948,2975){\makebox(0,0)[r]{\strut{}$1$}}%
      \put(3948,3350){\makebox(0,0)[r]{\strut{}$2$}}%
      \put(3948,3724){\makebox(0,0)[r]{\strut{}$3$}}%
      \put(3948,4099){\makebox(0,0)[r]{\strut{}$4$}}%
      \put(4192,2380){\makebox(0,0){\strut{}$2.9$}}%
      \put(4750,2380){\makebox(0,0){\strut{}$3$}}%
      \put(5308,2380){\makebox(0,0){\strut{}$3.1$}}%
      \put(5866,2380){\makebox(0,0){\strut{}$3.2$}}%
      \put(6424,2380){\makebox(0,0){\strut{}$3.3$}}%
      \put(6982,2380){\makebox(0,0){\strut{}$3.4$}}%
      \put(4602,3919){\makebox(0,0){\footnotesize $Q=2$GeV}}%
    }%
    \gplgaddtomacro\gplfronttext{%
    }%
    \gplgaddtomacro\gplbacktext{%
      \csname LTb\endcsname
      \put(348,440){\makebox(0,0)[r]{\strut{}$0$}}%
      \put(348,815){\makebox(0,0)[r]{\strut{}$1$}}%
      \put(348,1190){\makebox(0,0)[r]{\strut{}$2$}}%
      \put(348,1565){\makebox(0,0)[r]{\strut{}$3$}}%
      \put(348,1940){\makebox(0,0)[r]{\strut{}$4$}}%
      \put(592,220){\makebox(0,0){\strut{}$2.9$}}%
      \put(1150,220){\makebox(0,0){\strut{}$3$}}%
      \put(1708,220){\makebox(0,0){\strut{}$3.1$}}%
      \put(2266,220){\makebox(0,0){\strut{}$3.2$}}%
      \put(2825,220){\makebox(0,0){\strut{}$3.3$}}%
      \put(3383,220){\makebox(0,0){\strut{}$3.4$}}%
      \put(1003,1760){\makebox(0,0){\footnotesize $Q=3$GeV}}%
    }%
    \gplgaddtomacro\gplfronttext{%
      \csname LTb\endcsname
      \put(7,1190){\rotatebox{-270}{\makebox(0,0){\strut{}$\frac{\text{CP}^{\text{LO}+\text{NLO}}(\Delta \phi)}{\text{CP}^{\text{LO}}(\Delta\phi)}$}}}%
      \put(1931,-110){\makebox(0,0){\strut{}$\Delta \phi$}}%
    }%
    \gplgaddtomacro\gplbacktext{%
      \csname LTb\endcsname
      \put(3948,440){\makebox(0,0)[r]{\strut{}$0$}}%
      \put(3948,815){\makebox(0,0)[r]{\strut{}$1$}}%
      \put(3948,1190){\makebox(0,0)[r]{\strut{}$2$}}%
      \put(3948,1565){\makebox(0,0)[r]{\strut{}$3$}}%
      \put(3948,1940){\makebox(0,0)[r]{\strut{}$4$}}%
      \put(4192,220){\makebox(0,0){\strut{}$2.9$}}%
      \put(4750,220){\makebox(0,0){\strut{}$3$}}%
      \put(5308,220){\makebox(0,0){\strut{}$3.1$}}%
      \put(5866,220){\makebox(0,0){\strut{}$3.2$}}%
      \put(6424,220){\makebox(0,0){\strut{}$3.3$}}%
      \put(6982,220){\makebox(0,0){\strut{}$3.4$}}%
      \put(4602,1760){\makebox(0,0){\footnotesize $Q=4$GeV}}%
    }%
    \gplgaddtomacro\gplfronttext{%
      \csname LTb\endcsname
      \put(5531,-110){\makebox(0,0){\strut{}$\Delta \phi$}}%
    }%
    \gplbacktext
    \put(0,0){\includegraphics[width={360.00bp},height={216.00bp}]{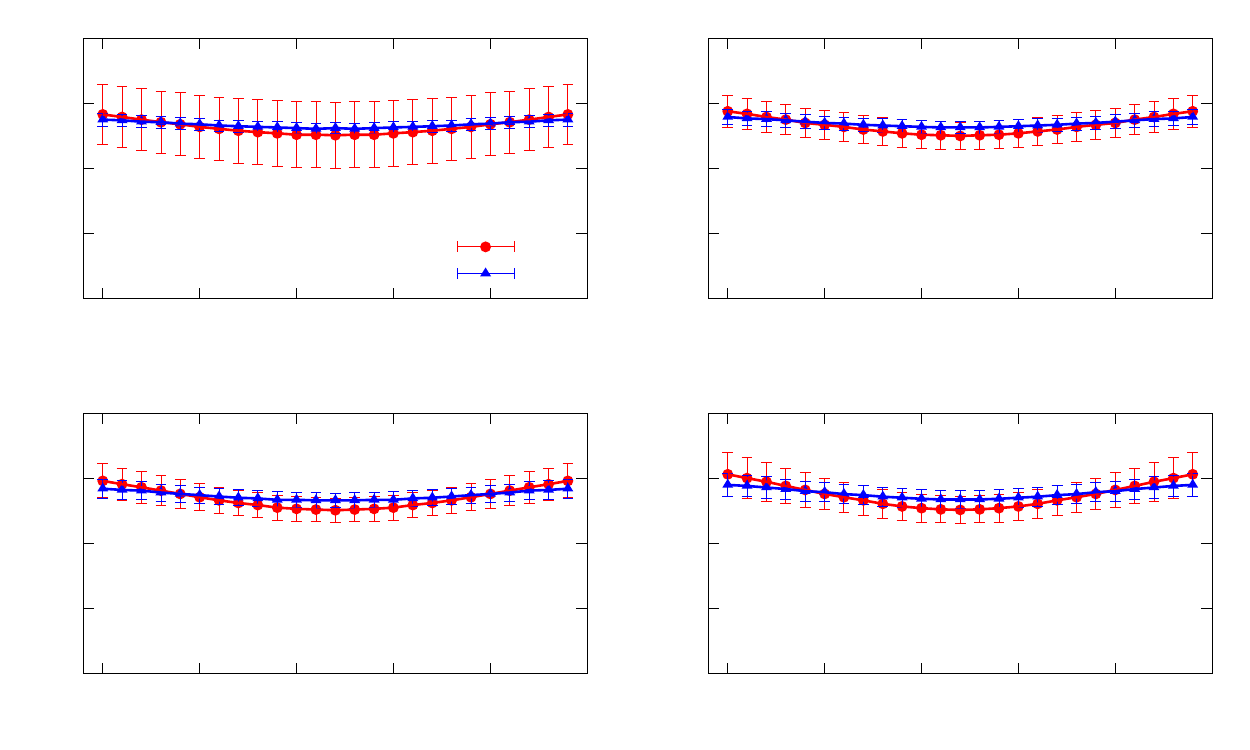}}%
    \gplfronttext
  \end{picture}%
\endgroup

%% file: CPplot2.tex
\begingroup
  \inputencoding{cp1252}%
  \makeatletter
  \providecommand\color[2][]{%
    \GenericError{(gnuplot) \space\space\space\@spaces}{%
      Package color not loaded in conjunction with
      terminal option `colourtext'%
    }{See the gnuplot documentation for explanation.%
    }{Either use 'blacktext' in gnuplot or load the package
      color.sty in LaTeX.}%
    \renewcommand\color[2][]{}%
  }%
  \providecommand\includegraphics[2][]{%
    \GenericError{(gnuplot) \space\space\space\@spaces}{%
      Package graphicx or graphics not loaded%
    }{See the gnuplot documentation for explanation.%
    }{The gnuplot epslatex terminal needs graphicx.sty or graphics.sty.}%
    \renewcommand\includegraphics[2][]{}%
  }%
  \providecommand\rotatebox[2]{#2}%
  \@ifundefined{ifGPcolor}{%
    \newif\ifGPcolor
    \GPcolorfalse
  }{}%
  \@ifundefined{ifGPblacktext}{%
    \newif\ifGPblacktext
    \GPblacktexttrue
  }{}%
  \let\gplgaddtomacro\g@addto@macro
  \gdef\gplbacktext{}%
  \gdef\gplfronttext{}%
  \makeatother
  \ifGPblacktext
    \def\colorrgb#1{}%
    \def\colorgray#1{}%
  \else
    \ifGPcolor
      \def\colorrgb#1{\color[rgb]{#1}}%
      \def\colorgray#1{\color[gray]{#1}}%
      \expandafter\def\csname LTw\endcsname{\color{white}}%
      \expandafter\def\csname LTb\endcsname{\color{black}}%
      \expandafter\def\csname LTa\endcsname{\color{black}}%
      \expandafter\def\csname LT0\endcsname{\color[rgb]{1,0,0}}%
      \expandafter\def\csname LT1\endcsname{\color[rgb]{0,1,0}}%
      \expandafter\def\csname LT2\endcsname{\color[rgb]{0,0,1}}%
      \expandafter\def\csname LT3\endcsname{\color[rgb]{1,0,1}}%
      \expandafter\def\csname LT4\endcsname{\color[rgb]{0,1,1}}%
      \expandafter\def\csname LT5\endcsname{\color[rgb]{1,1,0}}%
      \expandafter\def\csname LT6\endcsname{\color[rgb]{0,0,0}}%
      \expandafter\def\csname LT7\endcsname{\color[rgb]{1,0.3,0}}%
      \expandafter\def\csname LT8\endcsname{\color[rgb]{0.5,0.5,0.5}}%
    \else
      \def\colorrgb#1{\color{black}}%
      \def\colorgray#1{\color[gray]{#1}}%
      \expandafter\def\csname LTw\endcsname{\color{white}}%
      \expandafter\def\csname LTb\endcsname{\color{black}}%
      \expandafter\def\csname LTa\endcsname{\color{black}}%
      \expandafter\def\csname LT0\endcsname{\color{black}}%
      \expandafter\def\csname LT1\endcsname{\color{black}}%
      \expandafter\def\csname LT2\endcsname{\color{black}}%
      \expandafter\def\csname LT3\endcsname{\color{black}}%
      \expandafter\def\csname LT4\endcsname{\color{black}}%
      \expandafter\def\csname LT5\endcsname{\color{black}}%
      \expandafter\def\csname LT6\endcsname{\color{black}}%
      \expandafter\def\csname LT7\endcsname{\color{black}}%
      \expandafter\def\csname LT8\endcsname{\color{black}}%
    \fi
  \fi
    \setlength{\unitlength}{0.0500bp}%
    \ifx\gptboxheight\undefined%
      \newlength{\gptboxheight}%
      \newlength{\gptboxwidth}%
      \newsavebox{\gptboxtext}%
    \fi%
    \setlength{\fboxrule}{0.5pt}%
    \setlength{\fboxsep}{1pt}%
\begin{picture}(7200.00,4320.00)%
    \gplgaddtomacro\gplbacktext{%
      \csname LTb\endcsname
      \put(348,2600){\makebox(0,0)[r]{\strut{}$0$}}%
      \put(348,2975){\makebox(0,0)[r]{\strut{}$0.1$}}%
      \put(348,3350){\makebox(0,0)[r]{\strut{}$0.2$}}%
      \put(348,3724){\makebox(0,0)[r]{\strut{}$0.3$}}%
      \put(348,4099){\makebox(0,0)[r]{\strut{}$0.4$}}%
      \put(592,2380){\makebox(0,0){\strut{}$2.9$}}%
      \put(1150,2380){\makebox(0,0){\strut{}$3$}}%
      \put(1708,2380){\makebox(0,0){\strut{}$3.1$}}%
      \put(2266,2380){\makebox(0,0){\strut{}$3.2$}}%
      \put(2825,2380){\makebox(0,0){\strut{}$3.3$}}%
      \put(3383,2380){\makebox(0,0){\strut{}$3.4$}}%
      \put(1003,3919){\makebox(0,0){\footnotesize $Q=1$GeV}}%
    }%
    \gplgaddtomacro\gplfronttext{%
      \csname LTb\endcsname
      \put(-257,3349){\rotatebox{-270}{\makebox(0,0){\strut{}$\frac{\text{CP}^{A=200}(\Delta \phi)}{\text{CP}^{A=1}(\Delta\phi)}$}}}%
      \csname LTb\endcsname
      \put(2779,3883){\makebox(0,0)[r]{\strut{}\footnotesize{LO}}}%
      \csname LTb\endcsname
      \put(2779,3751){\makebox(0,0)[r]{\strut{}\footnotesize{LO+NLO}}}%
    }%
    \gplgaddtomacro\gplbacktext{%
      \csname LTb\endcsname
      \put(3948,2600){\makebox(0,0)[r]{\strut{}$0$}}%
      \put(3948,2975){\makebox(0,0)[r]{\strut{}$0.1$}}%
      \put(3948,3350){\makebox(0,0)[r]{\strut{}$0.2$}}%
      \put(3948,3724){\makebox(0,0)[r]{\strut{}$0.3$}}%
      \put(3948,4099){\makebox(0,0)[r]{\strut{}$0.4$}}%
      \put(4192,2380){\makebox(0,0){\strut{}$2.9$}}%
      \put(4750,2380){\makebox(0,0){\strut{}$3$}}%
      \put(5308,2380){\makebox(0,0){\strut{}$3.1$}}%
      \put(5866,2380){\makebox(0,0){\strut{}$3.2$}}%
      \put(6424,2380){\makebox(0,0){\strut{}$3.3$}}%
      \put(6982,2380){\makebox(0,0){\strut{}$3.4$}}%
      \put(4602,3919){\makebox(0,0){\footnotesize $Q=2$GeV}}%
    }%
    \gplgaddtomacro\gplfronttext{%
    }%
    \gplgaddtomacro\gplbacktext{%
      \csname LTb\endcsname
      \put(348,440){\makebox(0,0)[r]{\strut{}$0$}}%
      \put(348,815){\makebox(0,0)[r]{\strut{}$0.1$}}%
      \put(348,1190){\makebox(0,0)[r]{\strut{}$0.2$}}%
      \put(348,1565){\makebox(0,0)[r]{\strut{}$0.3$}}%
      \put(348,1940){\makebox(0,0)[r]{\strut{}$0.4$}}%
      \put(592,220){\makebox(0,0){\strut{}$2.9$}}%
      \put(1150,220){\makebox(0,0){\strut{}$3$}}%
      \put(1708,220){\makebox(0,0){\strut{}$3.1$}}%
      \put(2266,220){\makebox(0,0){\strut{}$3.2$}}%
      \put(2825,220){\makebox(0,0){\strut{}$3.3$}}%
      \put(3383,220){\makebox(0,0){\strut{}$3.4$}}%
      \put(1003,1760){\makebox(0,0){\footnotesize $Q=3$GeV}}%
    }%
    \gplgaddtomacro\gplfronttext{%
      \csname LTb\endcsname
      \put(-257,1190){\rotatebox{-270}{\makebox(0,0){\strut{}$\frac{\text{CP}^{A=200}(\Delta \phi)}{\text{CP}^{A=1}(\Delta\phi)}$}}}%
      \put(1931,-110){\makebox(0,0){\strut{}$\Delta \phi$}}%
    }%
    \gplgaddtomacro\gplbacktext{%
      \csname LTb\endcsname
      \put(3948,440){\makebox(0,0)[r]{\strut{}$0$}}%
      \put(3948,815){\makebox(0,0)[r]{\strut{}$0.1$}}%
      \put(3948,1190){\makebox(0,0)[r]{\strut{}$0.2$}}%
      \put(3948,1565){\makebox(0,0)[r]{\strut{}$0.3$}}%
      \put(3948,1940){\makebox(0,0)[r]{\strut{}$0.4$}}%
      \put(4192,220){\makebox(0,0){\strut{}$2.9$}}%
      \put(4750,220){\makebox(0,0){\strut{}$3$}}%
      \put(5308,220){\makebox(0,0){\strut{}$3.1$}}%
      \put(5866,220){\makebox(0,0){\strut{}$3.2$}}%
      \put(6424,220){\makebox(0,0){\strut{}$3.3$}}%
      \put(6982,220){\makebox(0,0){\strut{}$3.4$}}%
      \put(4602,1760){\makebox(0,0){\footnotesize $Q=4$GeV}}%
    }%
    \gplgaddtomacro\gplfronttext{%
      \csname LTb\endcsname
      \put(5531,-110){\makebox(0,0){\strut{}$\Delta \phi$}}%
    }%
    \gplbacktext
    \put(0,0){\includegraphics[width={360.00bp},height={216.00bp}]{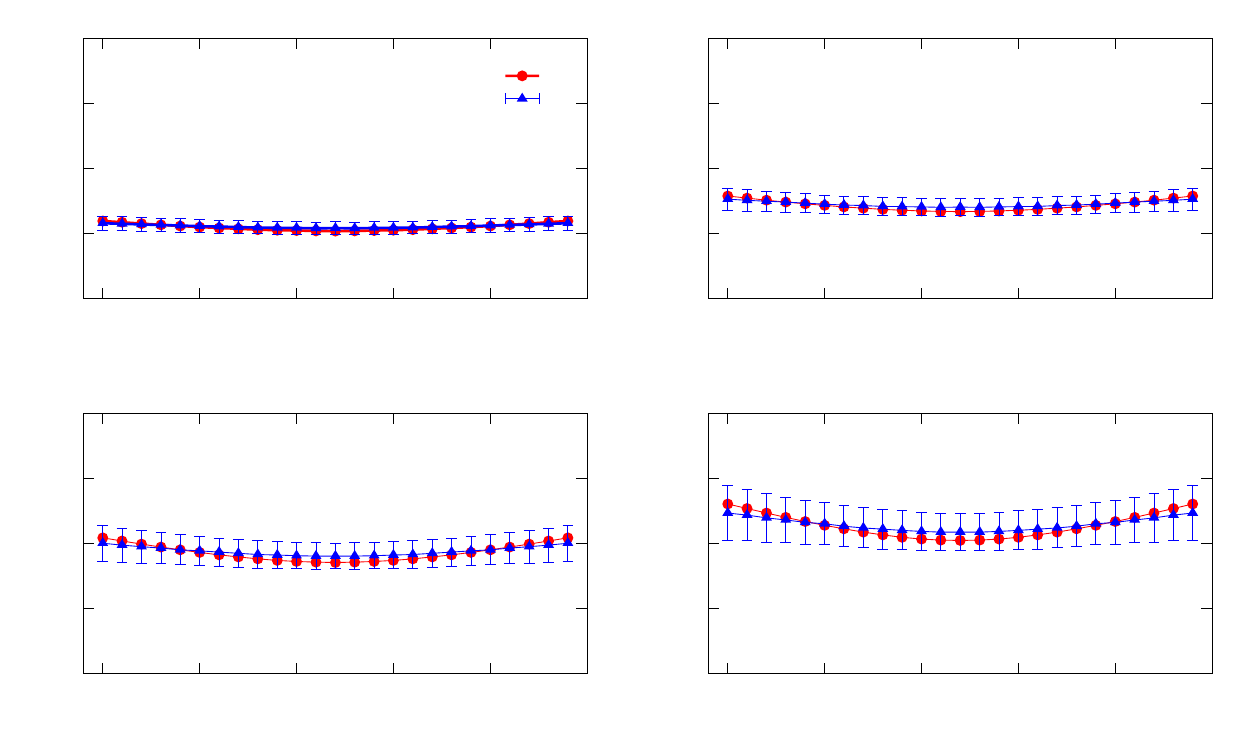}}%
    \gplfronttext
  \end{picture}%
\endgroup

%% file: 01.Eloss.bbl
\begin{thebibliography}{58}
\expandafter\ifx\csname natexlab\endcsname\relax\def\natexlab#1{#1}\fi
\expandafter\ifx\csname bibnamefont\endcsname\relax
  \def\bibnamefont#1{#1}\fi
\expandafter\ifx\csname bibfnamefont\endcsname\relax
  \def\bibfnamefont#1{#1}\fi
\expandafter\ifx\csname citenamefont\endcsname\relax
  \def\citenamefont#1{#1}\fi
\expandafter\ifx\csname url\endcsname\relax
  \def\url#1{\texttt{#1}}\fi
\expandafter\ifx\csname urlprefix\endcsname\relax\def\urlprefix{URL }\fi
\providecommand{\bibinfo}[2]{#2}
\providecommand{\eprint}[2][]{\url{#2}}

\bibitem[{\citenamefont{Aaron et~al.}(2009)}]{Aaron:2009kv}
\bibinfo{author}{\bibfnamefont{F.~D.} \bibnamefont{Aaron}} \bibnamefont{et~al.}
  (\bibinfo{collaboration}{H1}), \bibinfo{journal}{Eur. Phys. J. C}
  \textbf{\bibinfo{volume}{64}}, \bibinfo{pages}{561} (\bibinfo{year}{2009}),
  \eprint{0904.3513}.

\bibitem[{\citenamefont{Gribov et~al.}(1983)\citenamefont{Gribov, Levin, and
  Ryskin}}]{Gribov:1984tu}
\bibinfo{author}{\bibfnamefont{L.~V.} \bibnamefont{Gribov}},
  \bibinfo{author}{\bibfnamefont{E.~M.} \bibnamefont{Levin}}, \bibnamefont{and}
  \bibinfo{author}{\bibfnamefont{M.~G.} \bibnamefont{Ryskin}},
  \bibinfo{journal}{Phys. Rept.} \textbf{\bibinfo{volume}{100}},
  \bibinfo{pages}{1} (\bibinfo{year}{1983}).

\bibitem[{\citenamefont{Mueller and Qiu}(1986)}]{Mueller:1985wy}
\bibinfo{author}{\bibfnamefont{A.~H.} \bibnamefont{Mueller}} \bibnamefont{and}
  \bibinfo{author}{\bibfnamefont{J.-w.} \bibnamefont{Qiu}},
  \bibinfo{journal}{Nucl. Phys. B} \textbf{\bibinfo{volume}{268}},
  \bibinfo{pages}{427} (\bibinfo{year}{1986}).

\bibitem[{\citenamefont{Gelis et~al.}(2010)\citenamefont{Gelis, Iancu,
  Jalilian-Marian, and Venugopalan}}]{Gelis:2010nm}
\bibinfo{author}{\bibfnamefont{F.}~\bibnamefont{Gelis}},
  \bibinfo{author}{\bibfnamefont{E.}~\bibnamefont{Iancu}},
  \bibinfo{author}{\bibfnamefont{J.}~\bibnamefont{Jalilian-Marian}},
  \bibnamefont{and}
  \bibinfo{author}{\bibfnamefont{R.}~\bibnamefont{Venugopalan}},
  \bibinfo{journal}{Ann. Rev. Nucl. Part. Sci.} \textbf{\bibinfo{volume}{60}},
  \bibinfo{pages}{463} (\bibinfo{year}{2010}), \eprint{1002.0333}.

\bibitem[{\citenamefont{McLerran and Venugopalan}(1994)}]{McLerran:1993ni}
\bibinfo{author}{\bibfnamefont{L.~D.} \bibnamefont{McLerran}} \bibnamefont{and}
  \bibinfo{author}{\bibfnamefont{R.}~\bibnamefont{Venugopalan}},
  \bibinfo{journal}{Phys. Rev. D} \textbf{\bibinfo{volume}{49}},
  \bibinfo{pages}{2233} (\bibinfo{year}{1994}), \eprint{hep-ph/9309289}.

\bibitem[{\citenamefont{Jalilian-Marian
  et~al.}(1997{\natexlab{a}})\citenamefont{Jalilian-Marian, Kovner, McLerran,
  and Weigert}}]{Jalilian-Marian:1996mkd}
\bibinfo{author}{\bibfnamefont{J.}~\bibnamefont{Jalilian-Marian}},
  \bibinfo{author}{\bibfnamefont{A.}~\bibnamefont{Kovner}},
  \bibinfo{author}{\bibfnamefont{L.~D.} \bibnamefont{McLerran}},
  \bibnamefont{and} \bibinfo{author}{\bibfnamefont{H.}~\bibnamefont{Weigert}},
  \bibinfo{journal}{Phys. Rev. D} \textbf{\bibinfo{volume}{55}},
  \bibinfo{pages}{5414} (\bibinfo{year}{1997}{\natexlab{a}}),
  \eprint{hep-ph/9606337}.

\bibitem[{\citenamefont{Albacete et~al.}(2018)}]{Albacete:2017qng}
\bibinfo{author}{\bibfnamefont{J.~L.} \bibnamefont{Albacete}}
  \bibnamefont{et~al.}, \bibinfo{journal}{Nucl. Phys. A}
  \textbf{\bibinfo{volume}{972}}, \bibinfo{pages}{18} (\bibinfo{year}{2018}),
  \eprint{1707.09973}.

\bibitem[{\citenamefont{Aschenauer et~al.}(2016)}]{Aschenauer:2016our}
\bibinfo{author}{\bibfnamefont{E.-C.} \bibnamefont{Aschenauer}}
  \bibnamefont{et~al.} (\bibinfo{year}{2016}), \eprint{1602.03922}.

\bibitem[{\citenamefont{Accardi et~al.}(2016)}]{Accardi:2012qut}
\bibinfo{author}{\bibfnamefont{A.}~\bibnamefont{Accardi}} \bibnamefont{et~al.},
  \bibinfo{journal}{Eur. Phys. J. A} \textbf{\bibinfo{volume}{52}},
  \bibinfo{pages}{268} (\bibinfo{year}{2016}), \eprint{1212.1701}.

\bibitem[{\citenamefont{Jalilian-Marian and
  Kovchegov}(2004)}]{JalilianMarian:2004da}
\bibinfo{author}{\bibfnamefont{J.}~\bibnamefont{Jalilian-Marian}}
  \bibnamefont{and} \bibinfo{author}{\bibfnamefont{Y.~V.}
  \bibnamefont{Kovchegov}}, \bibinfo{journal}{Phys. Rev. D}
  \textbf{\bibinfo{volume}{70}}, \bibinfo{pages}{114017}
  (\bibinfo{year}{2004}), \bibinfo{note}{[Erratum: Phys.Rev.D 71, 079901
  (2005)]}, \eprint{hep-ph/0405266}.

\bibitem[{\citenamefont{Jalilian-Marian}(2006)}]{Jalilian-Marian:2005qbq}
\bibinfo{author}{\bibfnamefont{J.}~\bibnamefont{Jalilian-Marian}},
  \bibinfo{journal}{Nucl. Phys. A} \textbf{\bibinfo{volume}{770}},
  \bibinfo{pages}{210} (\bibinfo{year}{2006}), \eprint{hep-ph/0509338}.

\bibitem[{\citenamefont{Marquet}(2007)}]{Marquet:2007vb}
\bibinfo{author}{\bibfnamefont{C.}~\bibnamefont{Marquet}},
  \bibinfo{journal}{Nucl. Phys. A} \textbf{\bibinfo{volume}{796}},
  \bibinfo{pages}{41} (\bibinfo{year}{2007}), \eprint{0708.0231}.

\bibitem[{\citenamefont{Albacete and Marquet}(2010)}]{Albacete:2010pg}
\bibinfo{author}{\bibfnamefont{J.~L.} \bibnamefont{Albacete}} \bibnamefont{and}
  \bibinfo{author}{\bibfnamefont{C.}~\bibnamefont{Marquet}},
  \bibinfo{journal}{Phys. Rev. Lett.} \textbf{\bibinfo{volume}{105}},
  \bibinfo{pages}{162301} (\bibinfo{year}{2010}), \eprint{1005.4065}.

\bibitem[{\citenamefont{Stasto et~al.}(2012)\citenamefont{Stasto, Xiao, and
  Yuan}}]{Stasto:2011ru}
\bibinfo{author}{\bibfnamefont{A.}~\bibnamefont{Stasto}},
  \bibinfo{author}{\bibfnamefont{B.-W.} \bibnamefont{Xiao}}, \bibnamefont{and}
  \bibinfo{author}{\bibfnamefont{F.}~\bibnamefont{Yuan}},
  \bibinfo{journal}{Phys. Lett. B} \textbf{\bibinfo{volume}{716}},
  \bibinfo{pages}{430} (\bibinfo{year}{2012}), \eprint{1109.1817}.

\bibitem[{\citenamefont{Lappi and Mantysaari}(2013)}]{Lappi:2012nh}
\bibinfo{author}{\bibfnamefont{T.}~\bibnamefont{Lappi}} \bibnamefont{and}
  \bibinfo{author}{\bibfnamefont{H.}~\bibnamefont{Mantysaari}},
  \bibinfo{journal}{Nucl. Phys. A} \textbf{\bibinfo{volume}{908}},
  \bibinfo{pages}{51} (\bibinfo{year}{2013}), \eprint{1209.2853}.

\bibitem[{\citenamefont{Jalilian-Marian and
  Rezaeian}(2012{\natexlab{a}})}]{Jalilian-Marian:2012wwi}
\bibinfo{author}{\bibfnamefont{J.}~\bibnamefont{Jalilian-Marian}}
  \bibnamefont{and} \bibinfo{author}{\bibfnamefont{A.~H.}
  \bibnamefont{Rezaeian}}, \bibinfo{journal}{Phys. Rev. D}
  \textbf{\bibinfo{volume}{86}}, \bibinfo{pages}{034016}
  (\bibinfo{year}{2012}{\natexlab{a}}), \eprint{1204.1319}.

\bibitem[{\citenamefont{Jalilian-Marian and
  Rezaeian}(2012{\natexlab{b}})}]{Jalilian-Marian:2011tvq}
\bibinfo{author}{\bibfnamefont{J.}~\bibnamefont{Jalilian-Marian}}
  \bibnamefont{and} \bibinfo{author}{\bibfnamefont{A.~H.}
  \bibnamefont{Rezaeian}}, \bibinfo{journal}{Phys. Rev. D}
  \textbf{\bibinfo{volume}{85}}, \bibinfo{pages}{014017}
  (\bibinfo{year}{2012}{\natexlab{b}}), \eprint{1110.2810}.

\bibitem[{\citenamefont{Zheng et~al.}(2014)\citenamefont{Zheng, Aschenauer,
  Lee, and Xiao}}]{Zheng:2014vka}
\bibinfo{author}{\bibfnamefont{L.}~\bibnamefont{Zheng}},
  \bibinfo{author}{\bibfnamefont{E.~C.} \bibnamefont{Aschenauer}},
  \bibinfo{author}{\bibfnamefont{J.~H.} \bibnamefont{Lee}}, \bibnamefont{and}
  \bibinfo{author}{\bibfnamefont{B.-W.} \bibnamefont{Xiao}},
  \bibinfo{journal}{Phys. Rev. D} \textbf{\bibinfo{volume}{89}},
  \bibinfo{pages}{074037} (\bibinfo{year}{2014}), \eprint{1403.2413}.

\bibitem[{\citenamefont{Stasto et~al.}(2018)\citenamefont{Stasto, Wei, Xiao,
  and Yuan}}]{Stasto:2018rci}
\bibinfo{author}{\bibfnamefont{A.}~\bibnamefont{Stasto}},
  \bibinfo{author}{\bibfnamefont{S.-Y.} \bibnamefont{Wei}},
  \bibinfo{author}{\bibfnamefont{B.-W.} \bibnamefont{Xiao}}, \bibnamefont{and}
  \bibinfo{author}{\bibfnamefont{F.}~\bibnamefont{Yuan}},
  \bibinfo{journal}{Phys. Lett. B} \textbf{\bibinfo{volume}{784}},
  \bibinfo{pages}{301} (\bibinfo{year}{2018}), \eprint{1805.05712}.

\bibitem[{\citenamefont{Albacete et~al.}(2019)\citenamefont{Albacete,
  Giacalone, Marquet, and Matas}}]{Albacete:2018ruq}
\bibinfo{author}{\bibfnamefont{J.~L.} \bibnamefont{Albacete}},
  \bibinfo{author}{\bibfnamefont{G.}~\bibnamefont{Giacalone}},
  \bibinfo{author}{\bibfnamefont{C.}~\bibnamefont{Marquet}}, \bibnamefont{and}
  \bibinfo{author}{\bibfnamefont{M.}~\bibnamefont{Matas}},
  \bibinfo{journal}{Phys. Rev. D} \textbf{\bibinfo{volume}{99}},
  \bibinfo{pages}{014002} (\bibinfo{year}{2019}), \eprint{1805.05711}.

\bibitem[{\citenamefont{M\"antysaari et~al.}(2020)\citenamefont{M\"antysaari,
  Mueller, Salazar, and Schenke}}]{Mantysaari:2019hkq}
\bibinfo{author}{\bibfnamefont{H.}~\bibnamefont{M\"antysaari}},
  \bibinfo{author}{\bibfnamefont{N.}~\bibnamefont{Mueller}},
  \bibinfo{author}{\bibfnamefont{F.}~\bibnamefont{Salazar}}, \bibnamefont{and}
  \bibinfo{author}{\bibfnamefont{B.}~\bibnamefont{Schenke}},
  \bibinfo{journal}{Phys. Rev. Lett.} \textbf{\bibinfo{volume}{124}},
  \bibinfo{pages}{112301} (\bibinfo{year}{2020}), \eprint{1912.05586}.

\bibitem[{\citenamefont{Hatta et~al.}(2021)\citenamefont{Hatta, Xiao, Yuan, and
  Zhou}}]{Hatta:2020bgy}
\bibinfo{author}{\bibfnamefont{Y.}~\bibnamefont{Hatta}},
  \bibinfo{author}{\bibfnamefont{B.-W.} \bibnamefont{Xiao}},
  \bibinfo{author}{\bibfnamefont{F.}~\bibnamefont{Yuan}}, \bibnamefont{and}
  \bibinfo{author}{\bibfnamefont{J.}~\bibnamefont{Zhou}},
  \bibinfo{journal}{Phys. Rev. Lett.} \textbf{\bibinfo{volume}{126}},
  \bibinfo{pages}{142001} (\bibinfo{year}{2021}), \eprint{2010.10774}.

\bibitem[{\citenamefont{Jia et~al.}(2020)\citenamefont{Jia, Wei, Xiao, and
  Yuan}}]{Jia:2019qbl}
\bibinfo{author}{\bibfnamefont{J.}~\bibnamefont{Jia}},
  \bibinfo{author}{\bibfnamefont{S.-Y.} \bibnamefont{Wei}},
  \bibinfo{author}{\bibfnamefont{B.-W.} \bibnamefont{Xiao}}, \bibnamefont{and}
  \bibinfo{author}{\bibfnamefont{F.}~\bibnamefont{Yuan}},
  \bibinfo{journal}{Phys. Rev. D} \textbf{\bibinfo{volume}{101}},
  \bibinfo{pages}{094008} (\bibinfo{year}{2020}), \eprint{1910.05290}.

\bibitem[{\citenamefont{Braidot}(2011)}]{Braidot:2010ig}
\bibinfo{author}{\bibfnamefont{E.}~\bibnamefont{Braidot}}
  (\bibinfo{collaboration}{STAR}), \bibinfo{journal}{Nucl. Phys. A}
  \textbf{\bibinfo{volume}{854}}, \bibinfo{pages}{168} (\bibinfo{year}{2011}),
  \eprint{1008.3989}.

\bibitem[{\citenamefont{Adare et~al.}(2011)}]{Adare:2011sc}
\bibinfo{author}{\bibfnamefont{A.}~\bibnamefont{Adare}} \bibnamefont{et~al.}
  (\bibinfo{collaboration}{PHENIX}), \bibinfo{journal}{Phys. Rev. Lett.}
  \textbf{\bibinfo{volume}{107}}, \bibinfo{pages}{172301}
  (\bibinfo{year}{2011}), \eprint{1105.5112}.

\bibitem[{\citenamefont{Kang et~al.}(2012)\citenamefont{Kang, Vitev, and
  Xing}}]{Kang:2011bp}
\bibinfo{author}{\bibfnamefont{Z.-B.} \bibnamefont{Kang}},
  \bibinfo{author}{\bibfnamefont{I.}~\bibnamefont{Vitev}}, \bibnamefont{and}
  \bibinfo{author}{\bibfnamefont{H.}~\bibnamefont{Xing}},
  \bibinfo{journal}{Phys. Rev. D} \textbf{\bibinfo{volume}{85}},
  \bibinfo{pages}{054024} (\bibinfo{year}{2012}), \eprint{1112.6021}.

\bibitem[{\citenamefont{Bergabo and Jalilian-Marian}(2021)}]{nlo:fbjjm}
\bibinfo{author}{\bibfnamefont{F.}~\bibnamefont{Bergabo}} \bibnamefont{and}
  \bibinfo{author}{\bibfnamefont{J.}~\bibnamefont{Jalilian-Marian}}
  (\bibinfo{year}{2021}), \eprint{in progress}.

\bibitem[{\citenamefont{Caucal et~al.}(2021)\citenamefont{Caucal, Salazar, and
  Venugopalan}}]{Caucal:2021ent}
\bibinfo{author}{\bibfnamefont{P.}~\bibnamefont{Caucal}},
  \bibinfo{author}{\bibfnamefont{F.}~\bibnamefont{Salazar}}, \bibnamefont{and}
  \bibinfo{author}{\bibfnamefont{R.}~\bibnamefont{Venugopalan}}
  (\bibinfo{year}{2021}), \eprint{2108.06347}.

\bibitem[{\citenamefont{Ayala et~al.}(2016)\citenamefont{Ayala, Hentschinski,
  Jalilian-Marian, and Tejeda-Yeomans}}]{Ayala:2016lhd}
\bibinfo{author}{\bibfnamefont{A.}~\bibnamefont{Ayala}},
  \bibinfo{author}{\bibfnamefont{M.}~\bibnamefont{Hentschinski}},
  \bibinfo{author}{\bibfnamefont{J.}~\bibnamefont{Jalilian-Marian}},
  \bibnamefont{and} \bibinfo{author}{\bibfnamefont{M.~E.}
  \bibnamefont{Tejeda-Yeomans}}, \bibinfo{journal}{Phys. Lett. B}
  \textbf{\bibinfo{volume}{761}}, \bibinfo{pages}{229} (\bibinfo{year}{2016}),
  \eprint{1604.08526}.

\bibitem[{\citenamefont{Ayala et~al.}(2017)\citenamefont{Ayala, Hentschinski,
  Jalilian-Marian, and Tejeda-Yeomans}}]{Ayala:2017rmh}
\bibinfo{author}{\bibfnamefont{A.}~\bibnamefont{Ayala}},
  \bibinfo{author}{\bibfnamefont{M.}~\bibnamefont{Hentschinski}},
  \bibinfo{author}{\bibfnamefont{J.}~\bibnamefont{Jalilian-Marian}},
  \bibnamefont{and} \bibinfo{author}{\bibfnamefont{M.~E.}
  \bibnamefont{Tejeda-Yeomans}}, \bibinfo{journal}{Nucl. Phys. B}
  \textbf{\bibinfo{volume}{920}}, \bibinfo{pages}{232} (\bibinfo{year}{2017}),
  \eprint{1701.07143}.

\bibitem[{\citenamefont{Gelis and Jalilian-Marian}(2003)}]{Gelis:2002nn}
\bibinfo{author}{\bibfnamefont{F.}~\bibnamefont{Gelis}} \bibnamefont{and}
  \bibinfo{author}{\bibfnamefont{J.}~\bibnamefont{Jalilian-Marian}},
  \bibinfo{journal}{Phys. Rev. D} \textbf{\bibinfo{volume}{67}},
  \bibinfo{pages}{074019} (\bibinfo{year}{2003}), \eprint{hep-ph/0211363}.

\bibitem[{\citenamefont{Balitsky}(1996)}]{Balitsky:1995ub}
\bibinfo{author}{\bibfnamefont{I.}~\bibnamefont{Balitsky}},
  \bibinfo{journal}{Nucl. Phys. B} \textbf{\bibinfo{volume}{463}},
  \bibinfo{pages}{99} (\bibinfo{year}{1996}), \eprint{hep-ph/9509348}.

\bibitem[{\citenamefont{Kovchegov}(2000)}]{Kovchegov:1999ua}
\bibinfo{author}{\bibfnamefont{Y.~V.} \bibnamefont{Kovchegov}},
  \bibinfo{journal}{Phys. Rev. D} \textbf{\bibinfo{volume}{61}},
  \bibinfo{pages}{074018} (\bibinfo{year}{2000}), \eprint{hep-ph/9905214}.

\bibitem[{\citenamefont{Jalilian-Marian
  et~al.}(1997{\natexlab{b}})\citenamefont{Jalilian-Marian, Kovner, Leonidov,
  and Weigert}}]{Jalilian-Marian:1997qno}
\bibinfo{author}{\bibfnamefont{J.}~\bibnamefont{Jalilian-Marian}},
  \bibinfo{author}{\bibfnamefont{A.}~\bibnamefont{Kovner}},
  \bibinfo{author}{\bibfnamefont{A.}~\bibnamefont{Leonidov}}, \bibnamefont{and}
  \bibinfo{author}{\bibfnamefont{H.}~\bibnamefont{Weigert}},
  \bibinfo{journal}{Nucl. Phys. B} \textbf{\bibinfo{volume}{504}},
  \bibinfo{pages}{415} (\bibinfo{year}{1997}{\natexlab{b}}),
  \eprint{hep-ph/9701284}.

\bibitem[{\citenamefont{Jalilian-Marian
  et~al.}(1998{\natexlab{a}})\citenamefont{Jalilian-Marian, Kovner, Leonidov,
  and Weigert}}]{Jalilian-Marian:1997jhx}
\bibinfo{author}{\bibfnamefont{J.}~\bibnamefont{Jalilian-Marian}},
  \bibinfo{author}{\bibfnamefont{A.}~\bibnamefont{Kovner}},
  \bibinfo{author}{\bibfnamefont{A.}~\bibnamefont{Leonidov}}, \bibnamefont{and}
  \bibinfo{author}{\bibfnamefont{H.}~\bibnamefont{Weigert}},
  \bibinfo{journal}{Phys. Rev. D} \textbf{\bibinfo{volume}{59}},
  \bibinfo{pages}{014014} (\bibinfo{year}{1998}{\natexlab{a}}),
  \eprint{hep-ph/9706377}.

\bibitem[{\citenamefont{Jalilian-Marian
  et~al.}(1998{\natexlab{b}})\citenamefont{Jalilian-Marian, Kovner, and
  Weigert}}]{Jalilian-Marian:1997ubg}
\bibinfo{author}{\bibfnamefont{J.}~\bibnamefont{Jalilian-Marian}},
  \bibinfo{author}{\bibfnamefont{A.}~\bibnamefont{Kovner}}, \bibnamefont{and}
  \bibinfo{author}{\bibfnamefont{H.}~\bibnamefont{Weigert}},
  \bibinfo{journal}{Phys. Rev. D} \textbf{\bibinfo{volume}{59}},
  \bibinfo{pages}{014015} (\bibinfo{year}{1998}{\natexlab{b}}),
  \eprint{hep-ph/9709432}.

\bibitem[{\citenamefont{Kovner et~al.}(2000)\citenamefont{Kovner, Milhano, and
  Weigert}}]{Kovner:2000pt}
\bibinfo{author}{\bibfnamefont{A.}~\bibnamefont{Kovner}},
  \bibinfo{author}{\bibfnamefont{J.~G.} \bibnamefont{Milhano}},
  \bibnamefont{and} \bibinfo{author}{\bibfnamefont{H.}~\bibnamefont{Weigert}},
  \bibinfo{journal}{Phys. Rev. D} \textbf{\bibinfo{volume}{62}},
  \bibinfo{pages}{114005} (\bibinfo{year}{2000}), \eprint{hep-ph/0004014}.

\bibitem[{\citenamefont{Iancu et~al.}(2001)\citenamefont{Iancu, Leonidov, and
  McLerran}}]{Iancu:2000hn}
\bibinfo{author}{\bibfnamefont{E.}~\bibnamefont{Iancu}},
  \bibinfo{author}{\bibfnamefont{A.}~\bibnamefont{Leonidov}}, \bibnamefont{and}
  \bibinfo{author}{\bibfnamefont{L.~D.} \bibnamefont{McLerran}},
  \bibinfo{journal}{Nucl. Phys. A} \textbf{\bibinfo{volume}{692}},
  \bibinfo{pages}{583} (\bibinfo{year}{2001}), \eprint{hep-ph/0011241}.

\bibitem[{\citenamefont{Ferreiro et~al.}(2002)\citenamefont{Ferreiro, Iancu,
  Leonidov, and McLerran}}]{Ferreiro:2001qy}
\bibinfo{author}{\bibfnamefont{E.}~\bibnamefont{Ferreiro}},
  \bibinfo{author}{\bibfnamefont{E.}~\bibnamefont{Iancu}},
  \bibinfo{author}{\bibfnamefont{A.}~\bibnamefont{Leonidov}}, \bibnamefont{and}
  \bibinfo{author}{\bibfnamefont{L.}~\bibnamefont{McLerran}},
  \bibinfo{journal}{Nucl. Phys. A} \textbf{\bibinfo{volume}{703}},
  \bibinfo{pages}{489} (\bibinfo{year}{2002}), \eprint{hep-ph/0109115}.

\bibitem[{\citenamefont{Dumitru
  et~al.}(2011{\natexlab{a}})\citenamefont{Dumitru, Jalilian-Marian, Lappi,
  Schenke, and Venugopalan}}]{Dumitru:2011vk}
\bibinfo{author}{\bibfnamefont{A.}~\bibnamefont{Dumitru}},
  \bibinfo{author}{\bibfnamefont{J.}~\bibnamefont{Jalilian-Marian}},
  \bibinfo{author}{\bibfnamefont{T.}~\bibnamefont{Lappi}},
  \bibinfo{author}{\bibfnamefont{B.}~\bibnamefont{Schenke}}, \bibnamefont{and}
  \bibinfo{author}{\bibfnamefont{R.}~\bibnamefont{Venugopalan}},
  \bibinfo{journal}{Phys. Lett. B} \textbf{\bibinfo{volume}{706}},
  \bibinfo{pages}{219} (\bibinfo{year}{2011}{\natexlab{a}}),
  \eprint{1108.4764}.

\bibitem[{\citenamefont{Dumitru
  et~al.}(2011{\natexlab{b}})\citenamefont{Dumitru, Jalilian-Marian, and
  Petreska}}]{Dumitru:2011zz}
\bibinfo{author}{\bibfnamefont{A.}~\bibnamefont{Dumitru}},
  \bibinfo{author}{\bibfnamefont{J.}~\bibnamefont{Jalilian-Marian}},
  \bibnamefont{and} \bibinfo{author}{\bibfnamefont{E.}~\bibnamefont{Petreska}},
  \bibinfo{journal}{Phys. Rev. D} \textbf{\bibinfo{volume}{84}},
  \bibinfo{pages}{014018} (\bibinfo{year}{2011}{\natexlab{b}}),
  \eprint{1105.4155}.

\bibitem[{\citenamefont{Dumitru and Jalilian-Marian}(2010)}]{Dumitru:2010ak}
\bibinfo{author}{\bibfnamefont{A.}~\bibnamefont{Dumitru}} \bibnamefont{and}
  \bibinfo{author}{\bibfnamefont{J.}~\bibnamefont{Jalilian-Marian}},
  \bibinfo{journal}{Phys. Rev. D} \textbf{\bibinfo{volume}{82}},
  \bibinfo{pages}{074023} (\bibinfo{year}{2010}), \eprint{1008.0480}.

\bibitem[{\citenamefont{Munier et~al.}(2017)\citenamefont{Munier, Peign\'e, and
  Petreska}}]{Munier:2016oih}
\bibinfo{author}{\bibfnamefont{S.}~\bibnamefont{Munier}},
  \bibinfo{author}{\bibfnamefont{S.}~\bibnamefont{Peign\'e}}, \bibnamefont{and}
  \bibinfo{author}{\bibfnamefont{E.}~\bibnamefont{Petreska}},
  \bibinfo{journal}{Phys. Rev. D} \textbf{\bibinfo{volume}{95}},
  \bibinfo{pages}{014014} (\bibinfo{year}{2017}), \eprint{1603.01028}.

\bibitem[{\citenamefont{Blaizot et~al.}(2004)\citenamefont{Blaizot, Gelis, and
  Venugopalan}}]{Blaizot:2004wv}
\bibinfo{author}{\bibfnamefont{J.~P.} \bibnamefont{Blaizot}},
  \bibinfo{author}{\bibfnamefont{F.}~\bibnamefont{Gelis}}, \bibnamefont{and}
  \bibinfo{author}{\bibfnamefont{R.}~\bibnamefont{Venugopalan}},
  \bibinfo{journal}{Nucl. Phys. A} \textbf{\bibinfo{volume}{743}},
  \bibinfo{pages}{57} (\bibinfo{year}{2004}), \eprint{hep-ph/0402257}.

\bibitem[{\citenamefont{Dominguez
  et~al.}(2011{\natexlab{a}})\citenamefont{Dominguez, Marquet, Xiao, and
  Yuan}}]{Dominguez:2011wm}
\bibinfo{author}{\bibfnamefont{F.}~\bibnamefont{Dominguez}},
  \bibinfo{author}{\bibfnamefont{C.}~\bibnamefont{Marquet}},
  \bibinfo{author}{\bibfnamefont{B.-W.} \bibnamefont{Xiao}}, \bibnamefont{and}
  \bibinfo{author}{\bibfnamefont{F.}~\bibnamefont{Yuan}},
  \bibinfo{journal}{Phys. Rev. D} \textbf{\bibinfo{volume}{83}},
  \bibinfo{pages}{105005} (\bibinfo{year}{2011}{\natexlab{a}}),
  \eprint{1101.0715}.

\bibitem[{\citenamefont{Fukushima and Hidaka}(2017)}]{Fukushima:2017mko}
\bibinfo{author}{\bibfnamefont{K.}~\bibnamefont{Fukushima}} \bibnamefont{and}
  \bibinfo{author}{\bibfnamefont{Y.}~\bibnamefont{Hidaka}},
  \bibinfo{journal}{JHEP} \textbf{\bibinfo{volume}{11}}, \bibinfo{pages}{114}
  (\bibinfo{year}{2017}), \eprint{1708.03051}.

\bibitem[{\citenamefont{Lappi et~al.}(2020)\citenamefont{Lappi, M\"antysaari,
  and Ramnath}}]{Lappi:2020srm}
\bibinfo{author}{\bibfnamefont{T.}~\bibnamefont{Lappi}},
  \bibinfo{author}{\bibfnamefont{H.}~\bibnamefont{M\"antysaari}},
  \bibnamefont{and} \bibinfo{author}{\bibfnamefont{A.}~\bibnamefont{Ramnath}},
  \bibinfo{journal}{Phys. Rev. D} \textbf{\bibinfo{volume}{102}},
  \bibinfo{pages}{074027} (\bibinfo{year}{2020}), \eprint{2007.00751}.

\bibitem[{\citenamefont{Golec-Biernat and
  Wusthoff}(1998)}]{GolecBiernat:1998js}
\bibinfo{author}{\bibfnamefont{K.~J.} \bibnamefont{Golec-Biernat}}
  \bibnamefont{and} \bibinfo{author}{\bibfnamefont{M.}~\bibnamefont{Wusthoff}},
  \bibinfo{journal}{Phys. Rev. D} \textbf{\bibinfo{volume}{59}},
  \bibinfo{pages}{014017} (\bibinfo{year}{1998}), \eprint{hep-ph/9807513}.

\bibitem[{\citenamefont{Dominguez
  et~al.}(2011{\natexlab{b}})\citenamefont{Dominguez, Xiao, and
  Yuan}}]{Dominguez:2010xd}
\bibinfo{author}{\bibfnamefont{F.}~\bibnamefont{Dominguez}},
  \bibinfo{author}{\bibfnamefont{B.-W.} \bibnamefont{Xiao}}, \bibnamefont{and}
  \bibinfo{author}{\bibfnamefont{F.}~\bibnamefont{Yuan}},
  \bibinfo{journal}{Phys. Rev. Lett.} \textbf{\bibinfo{volume}{106}},
  \bibinfo{pages}{022301} (\bibinfo{year}{2011}{\natexlab{b}}),
  \eprint{1009.2141}.

\bibitem[{\citenamefont{Mueller et~al.}(2013)\citenamefont{Mueller, Xiao, and
  Yuan}}]{Mueller:2012uf}
\bibinfo{author}{\bibfnamefont{A.~H.} \bibnamefont{Mueller}},
  \bibinfo{author}{\bibfnamefont{B.-W.} \bibnamefont{Xiao}}, \bibnamefont{and}
  \bibinfo{author}{\bibfnamefont{F.}~\bibnamefont{Yuan}},
  \bibinfo{journal}{Phys. Rev. Lett.} \textbf{\bibinfo{volume}{110}},
  \bibinfo{pages}{082301} (\bibinfo{year}{2013}), \eprint{1210.5792}.

\bibitem[{\citenamefont{Altinoluk and Boussarie}(2019)}]{Altinoluk:2019wyu}
\bibinfo{author}{\bibfnamefont{T.}~\bibnamefont{Altinoluk}} \bibnamefont{and}
  \bibinfo{author}{\bibfnamefont{R.}~\bibnamefont{Boussarie}},
  \bibinfo{journal}{JHEP} \textbf{\bibinfo{volume}{10}}, \bibinfo{pages}{208}
  (\bibinfo{year}{2019}), \eprint{1902.07930}.

\bibitem[{\citenamefont{Boussarie et~al.}(2021)\citenamefont{Boussarie,
  M\"antysaari, Salazar, and Schenke}}]{Boussarie:2021lkb}
\bibinfo{author}{\bibfnamefont{R.}~\bibnamefont{Boussarie}},
  \bibinfo{author}{\bibfnamefont{H.}~\bibnamefont{M\"antysaari}},
  \bibinfo{author}{\bibfnamefont{F.}~\bibnamefont{Salazar}}, \bibnamefont{and}
  \bibinfo{author}{\bibfnamefont{B.}~\bibnamefont{Schenke}}
  (\bibinfo{year}{2021}), \eprint{2106.11301}.

\bibitem[{\citenamefont{Kotko et~al.}(2015)\citenamefont{Kotko, Kutak, Marquet,
  Petreska, Sapeta, and van Hameren}}]{Kotko:2015ura}
\bibinfo{author}{\bibfnamefont{P.}~\bibnamefont{Kotko}},
  \bibinfo{author}{\bibfnamefont{K.}~\bibnamefont{Kutak}},
  \bibinfo{author}{\bibfnamefont{C.}~\bibnamefont{Marquet}},
  \bibinfo{author}{\bibfnamefont{E.}~\bibnamefont{Petreska}},
  \bibinfo{author}{\bibfnamefont{S.}~\bibnamefont{Sapeta}}, \bibnamefont{and}
  \bibinfo{author}{\bibfnamefont{A.}~\bibnamefont{van Hameren}},
  \bibinfo{journal}{JHEP} \textbf{\bibinfo{volume}{09}}, \bibinfo{pages}{106}
  (\bibinfo{year}{2015}), \eprint{1503.03421}.

\bibitem[{\citenamefont{van Hameren et~al.}(2016)\citenamefont{van Hameren,
  Kotko, Kutak, Marquet, Petreska, and Sapeta}}]{vanHameren:2016ftb}
\bibinfo{author}{\bibfnamefont{A.}~\bibnamefont{van Hameren}},
  \bibinfo{author}{\bibfnamefont{P.}~\bibnamefont{Kotko}},
  \bibinfo{author}{\bibfnamefont{K.}~\bibnamefont{Kutak}},
  \bibinfo{author}{\bibfnamefont{C.}~\bibnamefont{Marquet}},
  \bibinfo{author}{\bibfnamefont{E.}~\bibnamefont{Petreska}}, \bibnamefont{and}
  \bibinfo{author}{\bibfnamefont{S.}~\bibnamefont{Sapeta}},
  \bibinfo{journal}{JHEP} \textbf{\bibinfo{volume}{12}}, \bibinfo{pages}{034}
  (\bibinfo{year}{2016}), \bibinfo{note}{[Erratum: JHEP 02, 158 (2019)]},
  \eprint{1607.03121}.

\bibitem[{\citenamefont{Altinoluk et~al.}(2019)\citenamefont{Altinoluk,
  Boussarie, and Kotko}}]{Altinoluk:2019fui}
\bibinfo{author}{\bibfnamefont{T.}~\bibnamefont{Altinoluk}},
  \bibinfo{author}{\bibfnamefont{R.}~\bibnamefont{Boussarie}},
  \bibnamefont{and} \bibinfo{author}{\bibfnamefont{P.}~\bibnamefont{Kotko}},
  \bibinfo{journal}{JHEP} \textbf{\bibinfo{volume}{05}}, \bibinfo{pages}{156}
  (\bibinfo{year}{2019}), \eprint{1901.01175}.

\bibitem[{\citenamefont{Boussarie and Mehtar-Tani}(2021)}]{Boussarie:2020vzf}
\bibinfo{author}{\bibfnamefont{R.}~\bibnamefont{Boussarie}} \bibnamefont{and}
  \bibinfo{author}{\bibfnamefont{Y.}~\bibnamefont{Mehtar-Tani}},
  \bibinfo{journal}{Phys. Rev. D} \textbf{\bibinfo{volume}{103}},
  \bibinfo{pages}{094012} (\bibinfo{year}{2021}), \eprint{2001.06449}.

\bibitem[{\citenamefont{Fujii et~al.}(2020)\citenamefont{Fujii, Marquet, and
  Watanabe}}]{Fujii:2020bkl}
\bibinfo{author}{\bibfnamefont{H.}~\bibnamefont{Fujii}},
  \bibinfo{author}{\bibfnamefont{C.}~\bibnamefont{Marquet}}, \bibnamefont{and}
  \bibinfo{author}{\bibfnamefont{K.}~\bibnamefont{Watanabe}},
  \bibinfo{journal}{JHEP} \textbf{\bibinfo{volume}{12}}, \bibinfo{pages}{181}
  (\bibinfo{year}{2020}), \eprint{2006.16279}.

\bibitem[{\citenamefont{Altinoluk et~al.}(2021)\citenamefont{Altinoluk,
  Marquet, and Taels}}]{Altinoluk:2021ygv}
\bibinfo{author}{\bibfnamefont{T.}~\bibnamefont{Altinoluk}},
  \bibinfo{author}{\bibfnamefont{C.}~\bibnamefont{Marquet}}, \bibnamefont{and}
  \bibinfo{author}{\bibfnamefont{P.}~\bibnamefont{Taels}},
  \bibinfo{journal}{JHEP} \textbf{\bibinfo{volume}{06}}, \bibinfo{pages}{085}
  (\bibinfo{year}{2021}), \eprint{2103.14495}.

\end{thebibliography}
